\documentclass[traditabstract]{aa}  
\usepackage{txfonts}
\usepackage{graphicx}
\usepackage{epsfig}
\usepackage{natbib}
\usepackage{url}
\usepackage{twoopt}
\usepackage{dcolumn}
\usepackage[colorlinks=true]{hyperref}
\hypersetup{
citecolor=blue}

\begin{document} 

\title{Temperature and density dependence of  line profiles  of sodium
  perturbed by helium\thanks{Opacity tables are only available at the CDS via anonymous ftp to cdsarc.u-strasbg.fr (130.79.128.5) }}

   \subtitle{}

   \author{N. F. Allard         \inst{1,2}\thanks{This paper is dedicated
       to the memory of France Allard, who initiated this work.}
     \and K. Myneni              \inst{3}
     \and J. N. Blakely    \inst{3}
     \and   G. Guillon           \inst{4}
}

   \institute{GEPI, Observatoire de Paris, PSL Research University, 
     UMR 8111, CNRS, Sorbonne Paris Cit\'e, 
     61, Avenue de l’Observatoire, F-75014 Paris, France\\
              \email{nicole.allard@obspm.fr}
         \and
         Institut d'Astrophysique de Paris,  UMR7095, CNRS, 
         Universit\'e Paris VI, 98bis Boulevard Arago, F-75014 PARIS, France \\
         \and
         U.S. Army DEVCOM, Aviation and Missile Center, Redstone Arsenal,
         AL 35898 \\
         \and
              Laboratoire Interdisciplinaire Carnot de Bourgogne, 
         UMR6303, CNRS, Universit\'e de Bourgogne-Franche-Comt\'e, 
         21078 Dijon Cedex, France     \\
}
  \date{Received 22 February 2023/ Accepted 16 April 2023  }
   
 
   \abstract
  { Ultracool stellar atmospheres  and hot exoplanets
  show absorption by alkali resonance lines severely broadened by
  collisions with neutral perturbers.  In the coolest and densest
  atmospheres, such as those of T dwarfs, Na I and K I broadened by
  molecular hydrogen and helium can come to dominate the entire
  optical spectrum.  The effects of Na-He collision broadening  are also
  central to understanding the opacity of cool DZ white dwarf stars.
  }
{   In order to
  be able to construct synthetic spectra of brown dwarfs and cool DZ
  white dwarfs, where helium density can reach several 10$^{21}$~cm$^{-3}$,
  Na-He line profiles of the resonance lines have been computed over a 
  wide range of densities and temperatures. }
{ Unified line  profiles that are valid from the core to the far 
  wings at high densities  are calculated in the semiclassical
  approach using up-to-date molecular data including 
  electronic spin-orbit coupling from the sodium atom.
 Far wings are extended to more than 4000~cm$^{-1}$ from the
  line center when the helium density can reach 10$^{21}$~cm$^{-3}$
  at 5000~K.} 
   {We present a comprehensive study  of Na-He collisional profiles
at high density, and temperatures  from 5000~K, which is the temperature
prevailing in the atmosphere of ultra-cool DZ white dwarf stars, down to 
1~K, which is the temperature in liquid helium clusters.
Collision broadening and shift parameters within the impact
approximation obtained in the semiclassical and quantum
     theory using our new accurate molecular data are presented.
}
   {}


   \authorrunning{N.~F.~Allard et al}

   \titlerunning{Temperature and density dependence of  line profiles  of sodium
  perturbed by helium}
   
   \maketitle
%

\section{Introduction}
The opacities of  Na and K play a crucial role in the
atmospheres of brown dwarfs and exoplanets.
The studies of observed L and T dwarfs by
\citet{liebert2000} and  \citet{burrows2001} 
clearly showed  the
importance of extended  wings of both  sodium and potassium doublets
 centered at 0.589~$\mu$m and 0.77~$\mu$m, respectively. 
They pointed out the need for more accurate
line profile calculations than Lorentzian profiles.
Clearly understanding the shape of these lines is essential 
to modeling the transport of radiation from the interior.
A first improvement  was made by \citet{burrows2003}
 using  multiconfiguration self-consistent
 field Hartree-Fock potentials in the theory
 of \citet{szudy1975} and \citet{szudy1996}.
In \citet{allard2003}, we  presented the first
 application of the absorption  profiles of 
 Na and K resonance lines perturbed by He and H$_2$
 using molecular potentials
 of \citet{pascale1983}
to describe the  alkali-He interaction  and of  \citet{rossi1985} for the alkali-H$_2$ interaction. 
The line profiles were included  as a source of opacity in
model atmospheres  and  synthetic spectra  using the \citet{allard2001}
 atmosphere  program  PHOENIX.
 The results were compared to previous models 
 and  demonstrated that these improvements 
are of fundamental importance for obtaining a better quantitative 
interpretation of the spectra.

 The far wings  play a crucial role for the continuum
 generated far from the line center. It is then necessary to update all 
the opacity tables provided in the past for the alkali perturbed by
helium and molecular hydrogen using the most recent existing \textit{ab initio}
 potentials.  In particular, using our new
potentials, we find that the blue far wing is significantly altered, 
including the position of the blue satellite, compared to
the results given in \citet{allard2003}, while the far red wing remains
the same. This paper is a continuation of \citet{allard2007a} and 
  \citet{allard2019} where new K--H$_2$ and Na--H$_2$ collisional profiles
  have been presented using accurate  \textit{ab initio} potentials.  These profiles
  are now used in many studies of brown dwarfs
  \citep[e.g.,][]{marley2017,oreshenko2020}
  and exoplanets \citep[e.g.,][]{phillips2020,changeat2020,freedman2021,chubb2021,nikolov2022,chubb2023}.
  
There are various theoretical approaches for treating the 
problem of atomic lines broadened by collisions with atomic perturbers.
Fully quantum-mechanical methods  have been developed which 
allow practical calculations within the one-perturber approximation 
\citep{julienne1986,mies1986,herman1978,sando1983}.
The one-perturber approximation
neglects the contribution of the core of the line, and  is only valid   
 in the wing and in the limit of
densities low enough such that multiple perturber effects  may be neglected.
However, at the high-density range  prevailing in the atmospheres of DZ
white dwarfs 
(up to  several $\times$~10$^{21}$~cm$^{-3}$), 
  multiple perturber effects must be included in the theoretical treatment.
An exact methodology for the quantum calculation at high densities,
as encountered in stellar atmospheres or laboratory plasmas, is not known. 
Approximate unified-theory methods to construct density-dependent line shapes
from one perturber spectra are available by 
\citet{fano1963}, \citet{jablonski1945,jablonski1948},
\citet{anderson1952,anderson1956},
\citet{baranger1958a,baranger1958b,baranger1958c}, and \citet{szudy1975}.

In dense plasmas, as in very cool DZ white dwarfs and liquid helium clusters, 
the possibility of several atoms interacting strongly is high, and the effects
play a role in the wavelength of the line center, such as in the shift of the line, as
well as the continuum generated far from the line center. 
For such a high-perturber density, the  collisional effects
should be treated by using the autocorrelation formalism in order
to take into account  simultaneous collisions with more than one perturbing
atom.
 The theory of spectral line shapes, especially the unified approach we have
developed and refined, makes possible models of stellar spectra 
that account
for the centers of spectral lines and their extreme wings in one consistent
treatment. Complete details and the derivation of the theory 
are given in~\citet{allard1999}. A rapid account of the theory is given
in Sect.~\ref{sec:theory}.
Using our semiclassical approach, we have 
computed the cross sections separately for the two fine-structure 
components of the $P$ state of sodium, namely $^{2}$P$_{1/2}$ 
(yielding the $D1$ line by a transition to ground state $^{2}$S$_{1/2}$) 
and $^{2}$P$_{3/2}$ ($D2$ line). Calculations were made on
the basis of adiabatic Na-He potentials from~\citet{angelo2012}, 
including atomic Na spin-orbit coupling along the lines described 
in~\cite{cohen1974}. For \textit{ab initio} calculations, it is the 
selection of basis states and the optimization techniques that may 
determine the final accuracy (Sect.~\ref{sec:potentials}).
In Sect.~\ref{sec:satellites} we illustrate the evolution of
the absorption spectra of Na-He collisional profiles for the 
densities and temperatures prevailing in atmosphere of brown 
and cool white dwarf stars and in laboratory experiments of 
helium doped with alkali metals. The calculations span the 
range 1~K to 5000~K. We compare the  obtained spectral line 
parameters to those calculated  in the quantum Baranger-Lindholm 
(BL) theory~\citep{baranger1958a} (Sect.~\ref{sec:param}).

\begin{figure}
\resizebox{0.46\textwidth}{!}{\includegraphics*{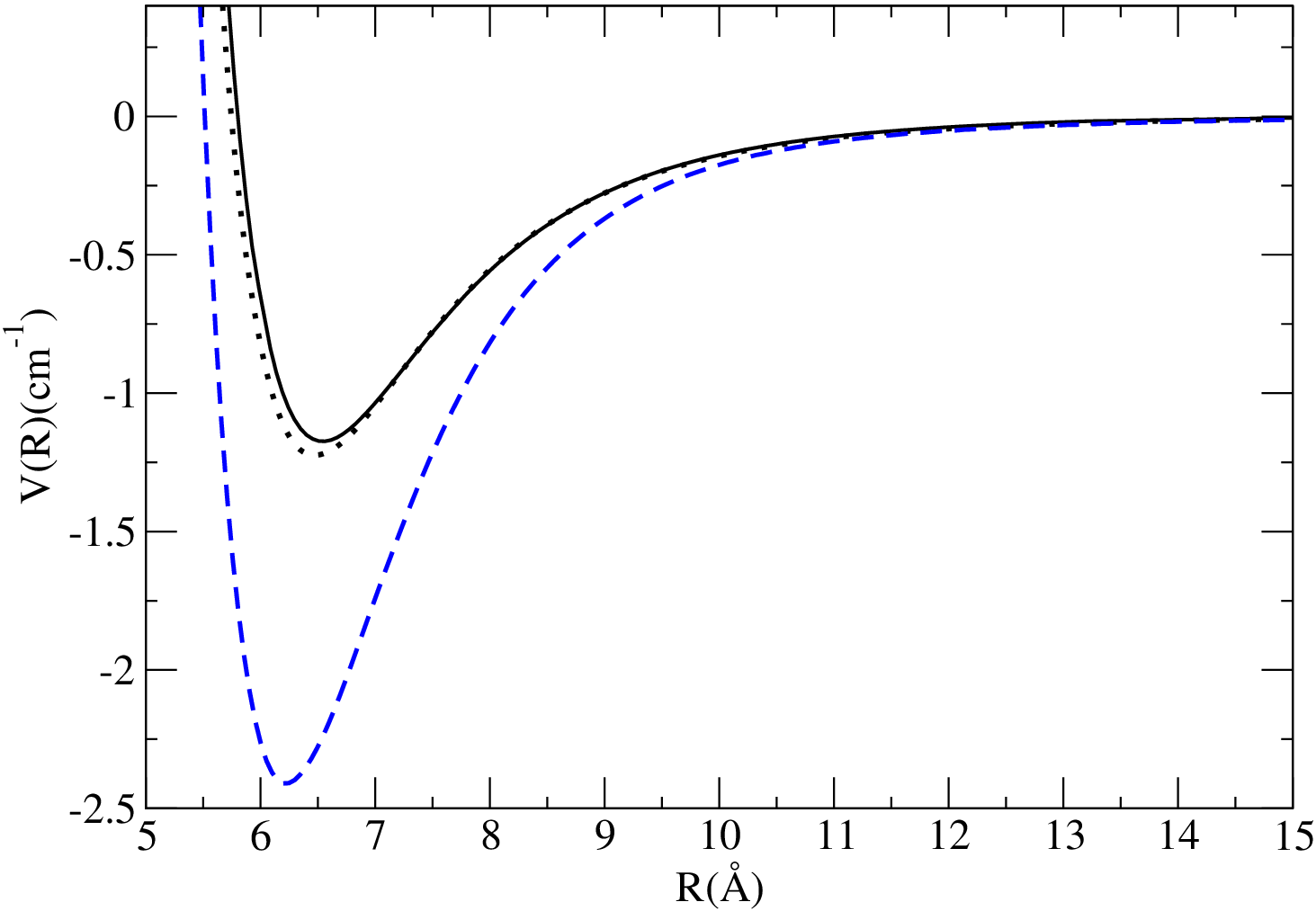}}
\caption  {Potential curve for the  3$s$ $X$ $^2\Sigma_{1/2}$ 
   state of Na-He~~\citet{angelo2012}.
The Na-He  potentials  of~\citet{nakayama2001b} (dotted line) and  
~\citet{pascale1983} (dashed blue line) are superimposed. 
 \label{potx}}
\end{figure}

\begin{figure}
\resizebox{0.46\textwidth}{!}{\includegraphics*{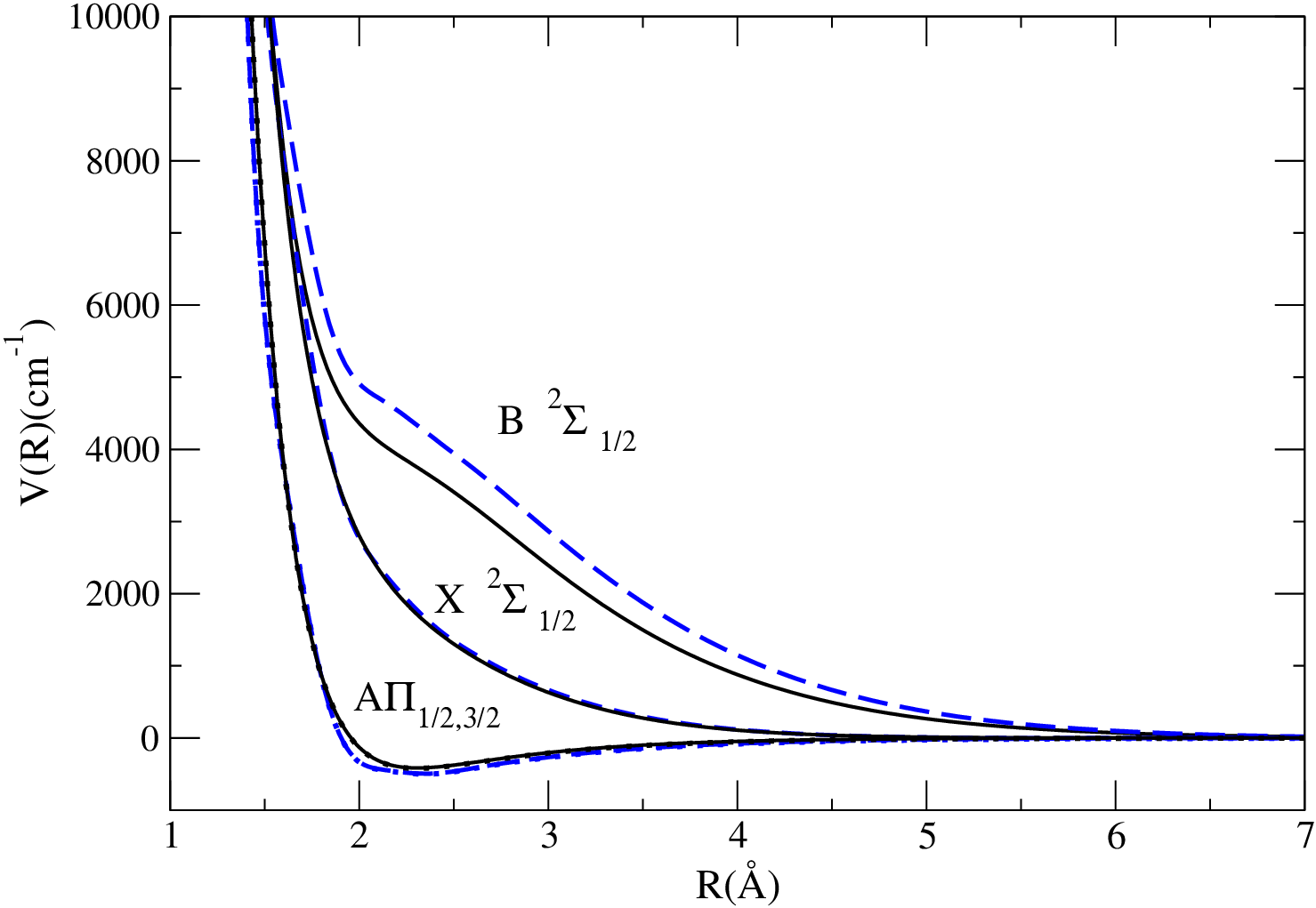}}
\caption  { Potential curves of the 
  Na-He molecule for  \textit{ab initio} potentials of
  \citet{angelo2012} (solid line) compared to pseudo-potentials of 
  \citet{pascale1983}  (dashed blue line).
\label{potB}}
\end{figure}

\begin{figure}
\resizebox{0.46\textwidth}{!}{\includegraphics*{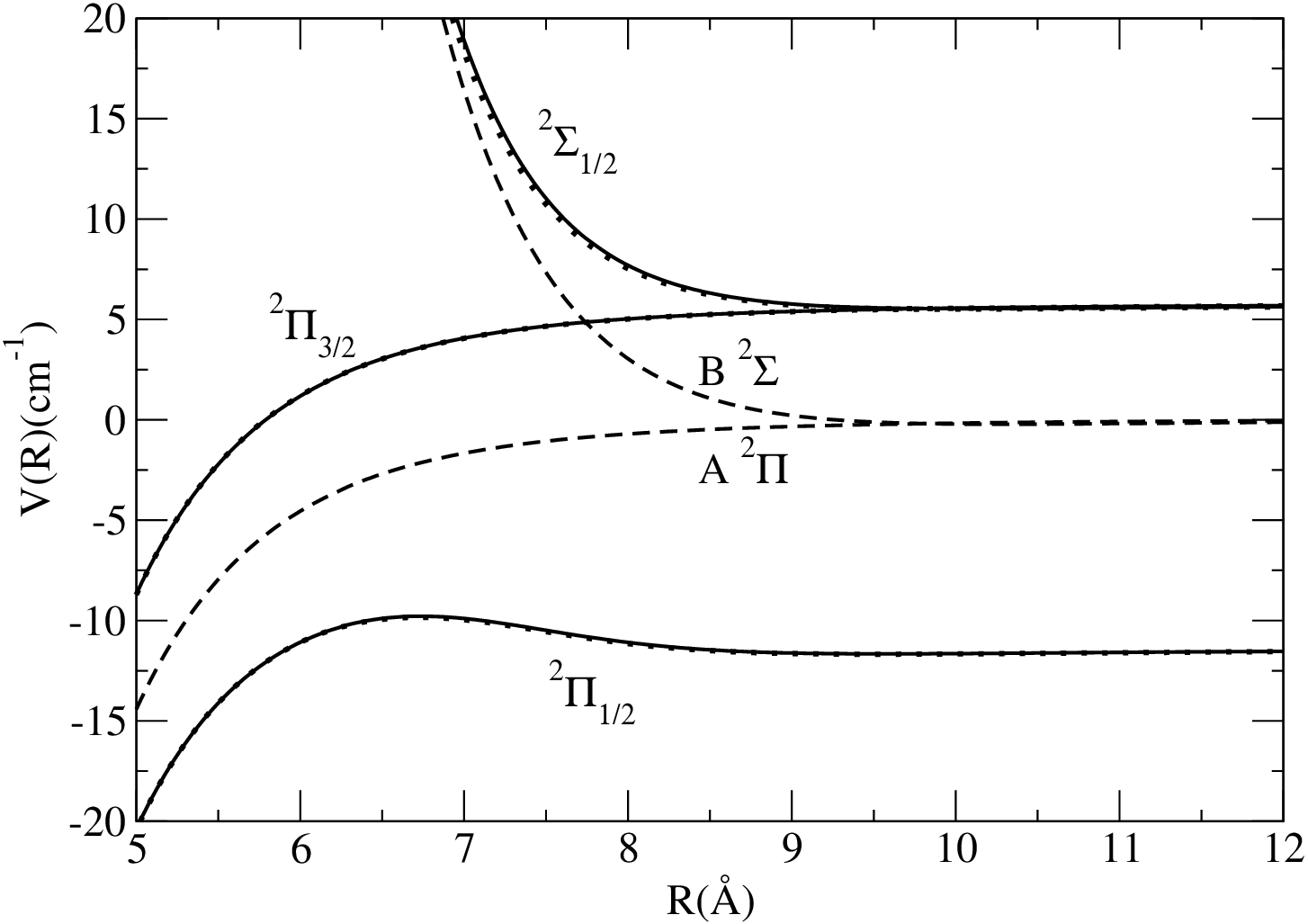}}
\caption{Details of the intermediate range of potential curves for
  the asymptotic Na $3p$ state with (solid line) and without (dashed line) 
spin-orbit coupling. The Na-He energies are given relative to the 
center of gravity of the $3p$ state from the \textit{ab initio}
  potentials calculated by~\citet{angelo2012}.
 The prior work of \citet{nakayama2001b}  is superimposed (dotted line).
 \label{pot3p}}
\end{figure}

\section{General expression for the spectrum in an adiabatic representation
\label{sec:theory}}

A unified theory of spectral line broadening 
has been developed to calculate neutral atom spectra given the
interaction and radiative transition moments for relevant states
of the radiating atom with other atoms in its environment.
Our approach is based on the quantum theory of
spectral line shapes of~\citet{baranger1958a,baranger1958b} 
 developed in an  adiabatic representation to include the degeneracy of atomic
levels \citep{royer1974,royer1980,allard1994}. 
 The spectrum $I(\Delta\omega)$ can be written as the 
Fourier transform (FT) of the dipole autocorrelation function $\Phi(s)$ ,
\begin{equation}
I(\Delta\omega)=
\frac{1}{\pi} \, Re \, \int^{+\infty}_0\Phi(s)e^{-i \Delta\omega s} ds,\label{eq:int}\end{equation}
where $s$ is time.

The FT in Eq.~(\ref{eq:int}) is taken such that $I(\Delta$ $\omega)$ is
normalized to unity when integrated over all frequencies, and
$\Delta$ $\omega$ is measured relative to the unperturbed line.
A pairwise additive assumption allows us to calculate the total profile
$I(\Delta \omega)$, when all the perturbers interact as the FT of the
$N^{\rm th}$ power of the autocorrelation function $\phi (s)$ of a
unique atom-perturber pair. Therefore,
\begin{equation}
\Phi(s)=(\phi(s))^{N}\; .
\end{equation}
That is to say, we neglect the interperturber correlations. The radiator can interact with several perturbers simultaneously, but the perturbers do
not interact with each other. It is what~\citet{royer1980} 
calls the {\it totally uncorrelated perturbers approximation}.
The fundamental result expressing the autocorrelation
function for many perturbers in terms
of a single perturber quantity $g(s)$ was first obtained
 by~\citet{anderson1952} and~\citet{baranger1958a} 
in the classical and quantum cases, 
respectively.
From the point of view of a general classical theory, the solution to
the ~\citet{anderson1952} model corresponds 
to the first-order approximation
in the gas density obtained by the cumulant expansion method~\citep{royer1972}.
The higher-order terms representing correlations 
between the perturbers  are neglected since they  are
 extremely complicated~\citep{royer1972,kubo1962,kubo1963,vankampen1974}.
 We  obtain for a perturber density $n_p$
\begin{equation}
  \Phi(s) = e^{-n_{p}g(s)}.
\label{eq:decay}  
\end{equation}
The decay of the autocorrelation function $\Phi (s)$ with time $s$ leads to
atomic line broadening. It depends on the density of perturbing atoms
$n_p$ and on their interaction with the radiating atom.
The molecular potentials and radiative dipole transition moments are input data
for a unified spectral line shape evaluation.
The dipole autocorrelation function $\Phi (s)$ is evaluated for
a classical collision path with an average over all possible collisions.
For a transition \mbox { $\alpha =(i,f)$} from an initial state~$i$ 
to a final state~$f$, we have 
\begin{eqnarray}
g_{\alpha}(s) && \,= \frac{1}
{\sum_{e,e'} \, \! ^{(\alpha)} \, |d_{ee'}|^2 }
\sum_{e,e'} \, \! ^{(\alpha)} \; \; \nonumber \\ 
&&  \int^{+\infty}_{0}\!\!2\pi b db
\int^{+\infty}_{-\infty}\!\! dx \; 
\tilde{d}_{ee'}[ \, R(0) \, ] \, \nonumber \\ 
&&[e^{\frac{i}{\hbar}\int^s_0 \!\! dt   \,
V_{e'e }[R(t)] } \,
\, \tilde{d^{*}}_{ee'}[R(s)] - \tilde{d}_{ee'}
[R(0)] \, ]. 
\label{eq:gcl}
\end{eqnarray}
In Eq.~(\ref{eq:gcl}), $e$ and  $e'$  label
the energy surfaces  on which the interacting
atoms  approach the initial and final atomic states of the transition 
 as \mbox{ $R \rightarrow \infty$ }.
 The sum $\sum_{e,e'} ^{(\alpha)}$ is
 over all pairs ($e,e'$)  such that
\mbox{$\omega_{e',e}(R) \rightarrow \omega_{\alpha}$} as 
\mbox{$R \rightarrow \infty$}.
We define 
$\tilde{d}_{ee'}(R(t))$ as a {\it modulated} dipole 
(Allard et al.~1999)
\begin{equation}
D(R) \equiv \tilde{d}_{ee'}[R(t)] = 
d_{ee'}[R(t)]e^{-\frac{\beta}{2}V_{e}[R(t)] } \; , \;
\label{eq:dip}
\end{equation}
where $\beta $ is the inverse temperature ($1/kT$).
Here $V_e$ is
the  ground state potential when we consider absorption profiles, 
or an excited state for the calculation of a profile in emission.
Over regions where $ V_e (r) <0 $, 
the factor $e^{-\beta V_e (r)}$ accounts for bound states of the
radiator-perturber pair, but in a classical approximation wherein the
discrete bound states are replaced by a continuum; thus any band
structure is smeared out.

In the present context, the perturbation of the frequency of the atomic
 transition during the collision results in a phase shift, 
$\eta(s)$, calculated along a classical path $R(t)$ assumed to be
rectilinear, is

\begin{equation}
\eta(s)=  \frac{1}{\hbar}\int^s_0 \, dt \;
V_{e'e }[ \, R(t) \, ]
\label{eq:phase}
\end{equation}
 where $\Delta V(R)$, the difference potential, is given by 
\begin{equation}
\Delta V(R) \equiv V_{e' e}[R(t)] = V_{e' }[R(t)] - V_{ e}[R(t)] \; ,
\end{equation}
 and it represents the difference between the electronic energies
of the quasi-molecular transition. The potential energy for a state $e$ is 
\begin{equation}
V_{e}[R(t)] = E_e[ R(t) ]-E_e^{\infty} \; .
\end{equation}

  At time $t$ from the point of closest approach 
\begin{equation}
R(t) = \left[\rho ^2 + (x+\bar{v} t)^2 \right]^{1/2} \; , \;
\end{equation}
with $\rho$ being the impact parameter of the perturber trajectory, 
$x$ the position of the perturber along its trajectory at time
$t=0$,  and $\bar{v}$ the average relative velocity between the
radiating atom and perturber. 

\section{Na-He diatomic potentials \label{sec:potentials}}

\citet{pascale1983}
obtained the diatomic adiabatic potentials for ground states and 
numerous excited states of alkali-metal--He systems by using  
$l$-dependent pseudo-potentials with parameters constrained by 
spectroscopic and scattering data. These potentials, and 
subsequent ones for other rare gases and alkalis computed similarly, 
were the prevailing theoretical basis to interpret and model spectral 
line profiles at the time and remain in use today because they can 
be flexibly adapted to measurements. However, with vast improvements 
in computational power  and refinements in code and methodology, 
significant progress in  \textit{ab initio} calculations of Na-He 
interactions has been achieved as reported by \citet{theodora1987}, 
\citet{nakayama2001b}, \citet{enomoto2004}, \citet{mullamphy2007},
and \citet{allouche2009}. Recent work of \citet{angelo2012} has been used for this paper. 

We computed the ground state ($X$ $^2\Sigma$) 
and the lowest first two excited states' ($A$ $^2\Pi$ and $B$ $^2\Sigma$) 
adiabatic potential energy curves (PECs) for the Na-He atom pair, 
correlating to Na($3s \; ^2\text{S}$) + He($^1\text{S}$) 
and Na($3p \; ^2\text{P}$) + He($^1\text{S}$), respectively, using the state-averaged 
complete active space self-consistent field 
multireference configuration interaction  (SA-CASSCF-MRCI+Q) high
level  \textit{ab initio} method, 
with additional Davidson cluster correction (Q) for a lack of size consistency. 
Complexation energies were obtained within the supermolecule approach. 
Eleven (nine from Na plus two from He) electrons were dynamically included in 
the calculation, leaving only one inner $1s$ frozen atomic orbital for sodium. 
We used extensive correlation-consistent basis sets of Dunning's type, 
aug-cc-pV6Z for He and cc-pV5Z for Na, whose accuracy have been checked against 
known atomic energies (the absolute energy for ground state He and 
the relative position of levels $3s \; ^2\text{S}$ and $3p \; ^2\text{P}$ for Na).
In particular, we have obtained a total ground state energy for 
He($^1\text{S}$)  within 0.3\% of the 
best available ones as of yet \citep{thakkar1994,freund1984,levin1985,drake2002}.

The active space used in the CASSCF method has been chosen to provide
the highest possible accuracy to the excited states' electronic energies,
most sensitive to it,
and it includes molecular orbitals of both $\sigma$ and $\pi$ types.
The accuracy of the MRCI ground $X$ $^2\Sigma$ PEC has been checked against
(size-extensive)
coupled cluster calculations up to triple excitation corrections, CCSD(T),
using the same basis sets.
Absolute energies obtained by these two methods for the whole range of
distances do not differ by more than 0.001\%,
while complexation energies are essentially indistinguishable.

Electronic energy grid points have been obtained for 220 internuclear
distances, ranging from $0.5$ a$_0$ to 1000 a$_0$,
with a higher density at short range and around the potential well.
We have paid particular attention to give a correct description at short
bond lengths,
as energies in this domain change very sharply, especially for the ground
state, and
line wing intensities are extremely dependent on them.   
On the other hand, the long range  \textit{ab initio} data are easily connected
to a simple inverse even power expansion series form with a leading term
in $R^{-6}$,
expressing dispersion for a neutral-neutral interaction.
More computational details can be found in \citet{angelo2012}.

Finally, the spin-orbit (SO) coupling within the Na-He dimer has been
included, while not having been computed \textit{ab initio}
from Breit-Pauli SO terms in the Hamiltonian, nor even in the Douglas-Kroll
approximation, alongside electronic energies.
Instead, given the still small value of the SO splitting for atomic
Na (roughly, $\Delta_{\text{SO}} =17.2$ cm$^{-1}$), 
SO mixing of adiabatic PECs has been incorporated from it, assuming a
smooth and slow variation as the He atom approaches, 
along the lines described in \citet{cohen1974},
\citet{Schneider1974}, and \citet{nakayama2001b}.
This procedure seems entirely reasonable, as shown by the rather
acceptable agreement between the
results obtained for the KHe dimer
(for K alone, $\Delta_{\text{SO}} = 57.7$ cm$^{-1}$) 
with the SO matrix explicitly computed within a full CI procedure, 
\citep{chattopadhyay2011}, and those using the asymptotic atomic
alkali value mentioned above \citep{nakayama2001b}.
All calculations have been performed with the MOLPRO
package \citep{molpro2006}. 

\begin{figure}
 \centering
\resizebox{0.46\textwidth}{!}
{\includegraphics*{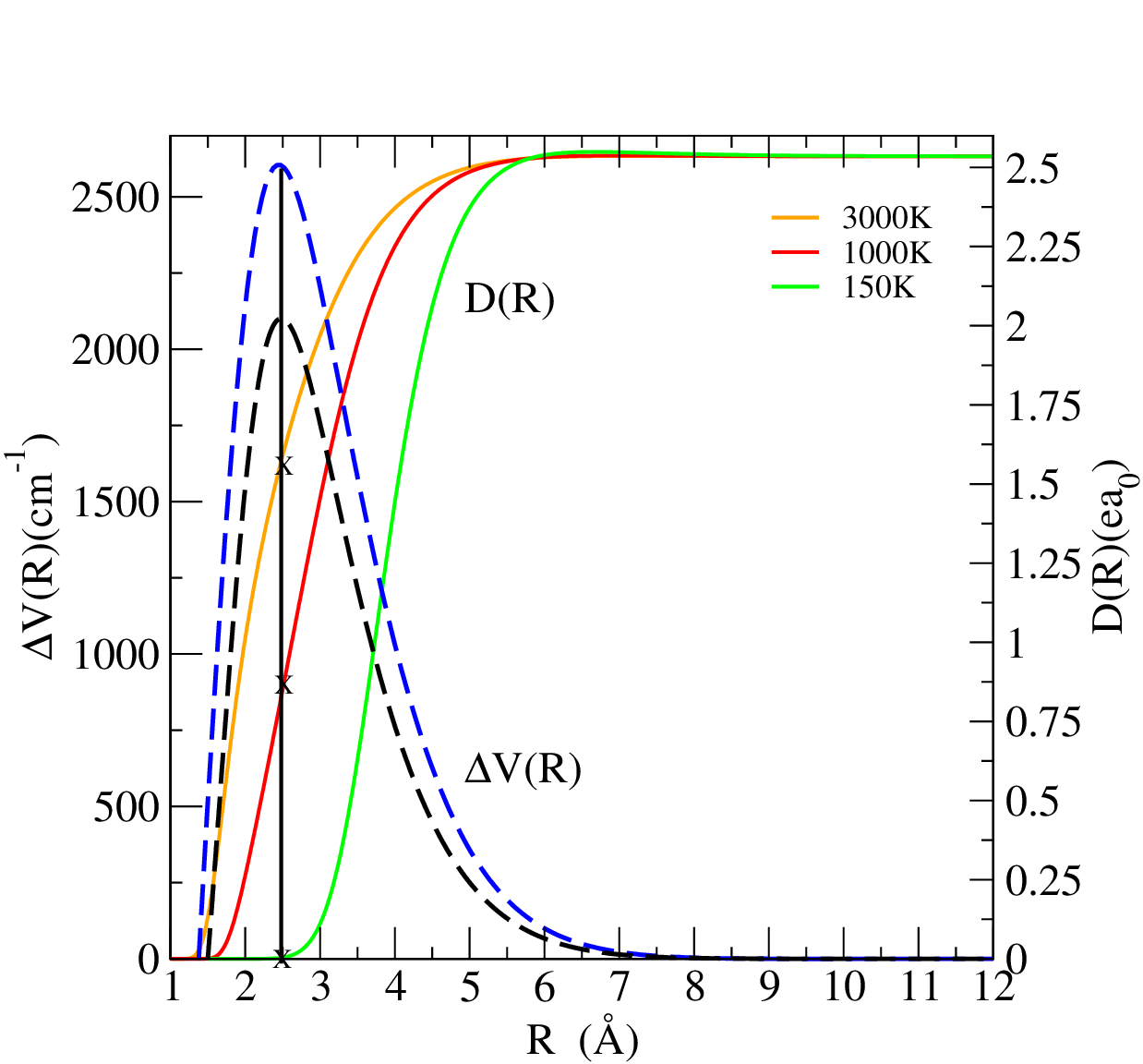}}
\caption  {$\Delta V(R)$ \citet{angelo2012} (dashed black line) compared to 
\citet{pascale1983} (dashed blue line) and the temperature dependence of 
modulated dipole $D(R)$ (solid line)
 corresponding to the 
  \mbox {3$s$ $X$    $\rightarrow$ 3$p$ $B$   $P_{3/2}$} transition
of the  Na $D2$ line. }
\label{diffpotNaHe}
\end{figure}

\begin{figure}
 \centering
\resizebox{0.46\textwidth}{!}
{\includegraphics*{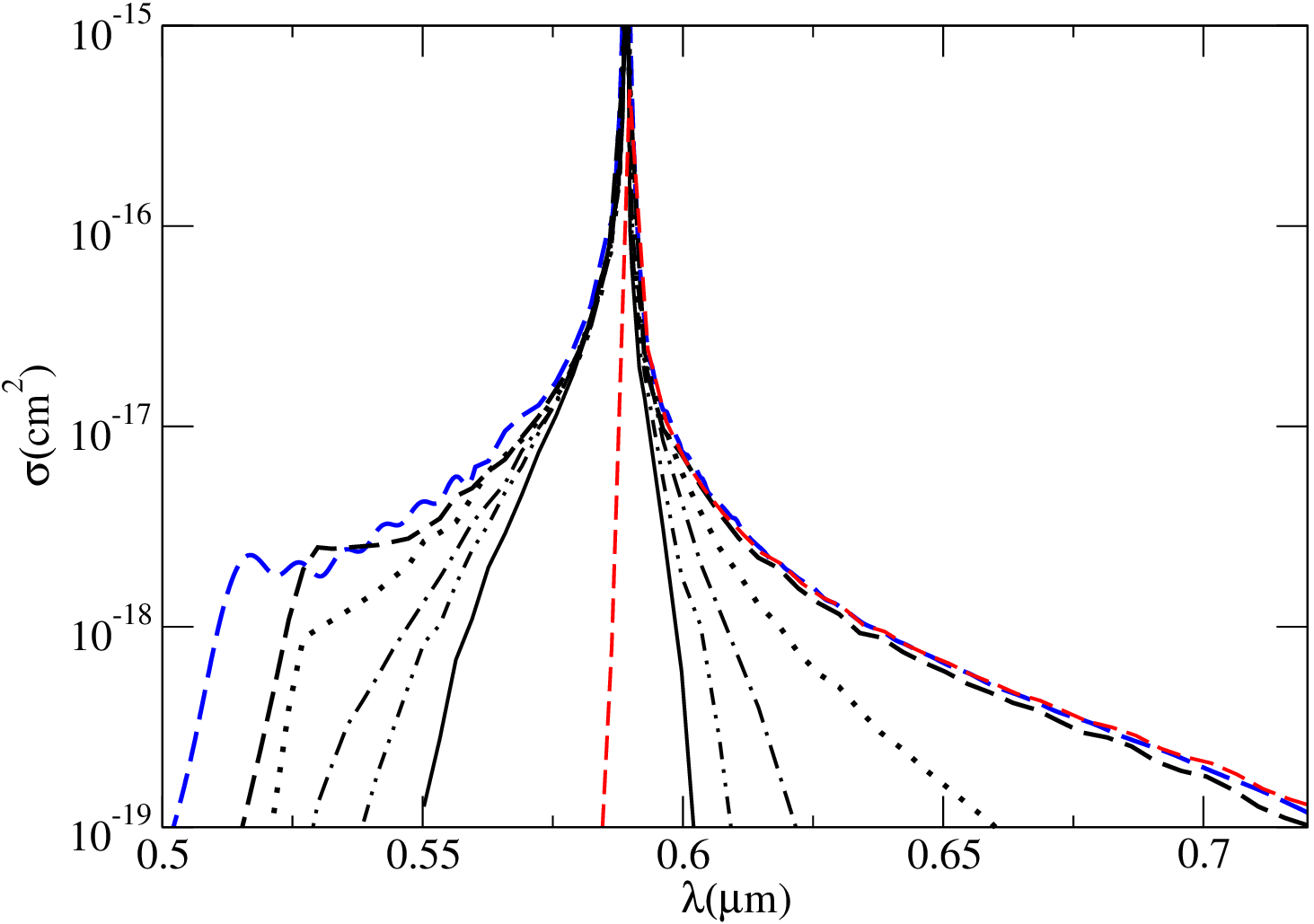}}
\caption  {Variation of the  absorption cross section 
of  the  $D2$  component  with temperature
(from top to bottom: \mbox{$T$  =3000, 1000, 400, 250, and 150~K}, and 
$n_{\rm He}$=10$^{20}$ cm$^{-3}$). The corresponding  profile for T=3000~K using
pseudo-potentials of \citet{pascale1983} is overplotted (dashed blue line).
 The  $D1$ component  for T=3000~K using   \textit{ab initio} potentials  of
\citet{angelo2012} is also overplotted (dashed red line).}
\label{varT}
\end{figure}

Figures~\ref{potx}-~\ref{potB} compare the potential data
of \citet{angelo2012} to  those 
of \citet{pascale1983} (dashed blue line).
The main difference   occurs at short distances in the 
repulsive walls of the $B$  potential; we can see the effect on 
the potential differences in Fig.~\ref{diffpotNaHe} and, consequently, on 
the blue wing of this component (Fig.~\ref{varT}). 
 The potential curves of the $A$~$^2\Pi_{1/2,3/2}$ states are very close
  and the red wings remain unchanged. Moreover  the fine structure
is small and  does not affect the extension of the red wings.
For the sake of clarity, we have further  reported, in Fig.~\ref{varT},
only the red wing of the $D1$
  component obtained with the potential data of  \citet{angelo2012}.

The potentials  computed by \cite{nakayama2001b}
 with a smaller basis set are overplotted (dotted line) 
in Figs.~\ref{potx} and \ref{pot3p}. 
They   are in excellent agreement for the 
$X$$^2\Sigma^+$ and $A$$^2\Pi$ states, which are most relevant in the
study of Na attached 
to He nanodroplets.  The barrier in the $^2\Pi_{1/2}$ that is influential 
on the line profile is exactly identical in the two approaches, 
as shown in Fig.~~\ref{pot3p}.

\section{Temperature and density dependence of the Na-He line profiles
\label{sec:satellites}}
 
The spectra  of  alkali metal atoms  perturbed  by rare gases
  have  been   extensively  investigated   in   experimental  and
  theoretical work by~\citet{allard1982}.
In \citet{allard2003}, we  presented  absorption  profiles of 
 Na  resonance lines perturbed by He using pseudo-potentials
 of \citet{pascale1983} to describe the interaction.
 At low densities, the 
binary model, for an optically active atom in collision with one perturber,
is valid for the whole profile, except for the central part of the line.
Several quantum mechanical calculations 
of absorption coefficients for far line wing spectra of sodium resonance 
lines broadened by helium were carried out by \citet{zhu2006} and
 by \citet{alioua2008} using  \citet{theodora1987} and
 \citet{allouche2009}  \textit{ab initio} potentials, respectively. 
  More recently, \citet{kielkopf2017} reported on  work  that
  compares unified line shape calculations based on \textit{ab initio}
   potentials with  experiments to determine 
the wings of the sodium and potassium resonance lines broadened by
H$_2$, He, and other rare gases.

\subsection{High temperatures}

Two excited molecular levels
 \mbox{ $B$ $^2\Sigma_{1/2}$} and \mbox {$A$ $^2\Pi_{3/2}$}
contribute to the  $D$2 line and produce opposite wings. 
 While $A$ states radiate on the red wing, $B$ states radiate
 on the blue wing. We subsequently restrict our analysis to the blue wing of the  
$D2$ line as the red wings of the doublet are unchanged compared to 
our previous results reported in \citet{allard2003}.
The unified theory \citep{anderson1952,allard1978,royer1978} 
predicts that there will
be line  satellites  centered periodically at 
frequencies corresponding to integer multiples of
the extrema of the difference potential $\Delta V(R)$:
\begin{equation}
\Delta V(R) \equiv V_{e' e}[R(t)] = V_{e' }[R(t)] - V_{ e}[R(t)] \; ,
\end{equation}
which represents the difference between the electronic energies
of a quasi-molecular $e$- $e'$ transition \citep{allard1982}.
The difference potential maximum as shown in Fig.~\ref{diffpotNaHe}  is 2600~cm$^{-1}$ when  using pseudo-potentials of \citet{pascale1983} and 2100~cm$^{-1}$ when using the  \textit{ab initio} potentials of \citet{angelo2012} and \citet{nakayama2001b}. They give rise to satellites positioned at $\Delta \hbar\omega$=$\Delta V_{max}$ \citep{allard1978}.
Because of the large maximum in $\Delta V$, the line satellite
 is  very well separated from the main line in the blue  wing. 
The presence of line satellite features is very sensitive to the 
temperature due to the fast variation of the  modulated dipole moment 
 $D(R$) (Eq.~(\ref{eq:dip})) with temperature in
the   internuclear  region   where  the   line  satellite   is  formed
(Fig~\ref{diffpotNaHe}). 
In Fig.~\ref{varT} we show
the absorption cross section for the resonance line of Na  for a
He density of  10$^{20}$ cm$^{-3}$ and temperatures from 150 to 3000~K, 
constituting the temperature range of brown dwarfs. The density and temperature
range used in Fig.~\ref{varT} span a pressure range of 2.1--41.4 bar.
The NaHe line satellite is apparent for  T $\geq$ 1000~K; 
 it disappears for  decreasing $T$ when
the transition moment  $D(R)$   gets very small. 
There is  formation of a   NaHe 
quasi-molecule at 0.53~$\mu$m, which is in agreement with theoretical results obtained 
by \citet{zhu2006}, \citet{kielkopf2008b}, and \citet{alioua2008}. 

\begin{figure}
 \centering
\resizebox{0.46\textwidth}{!}
{\includegraphics*{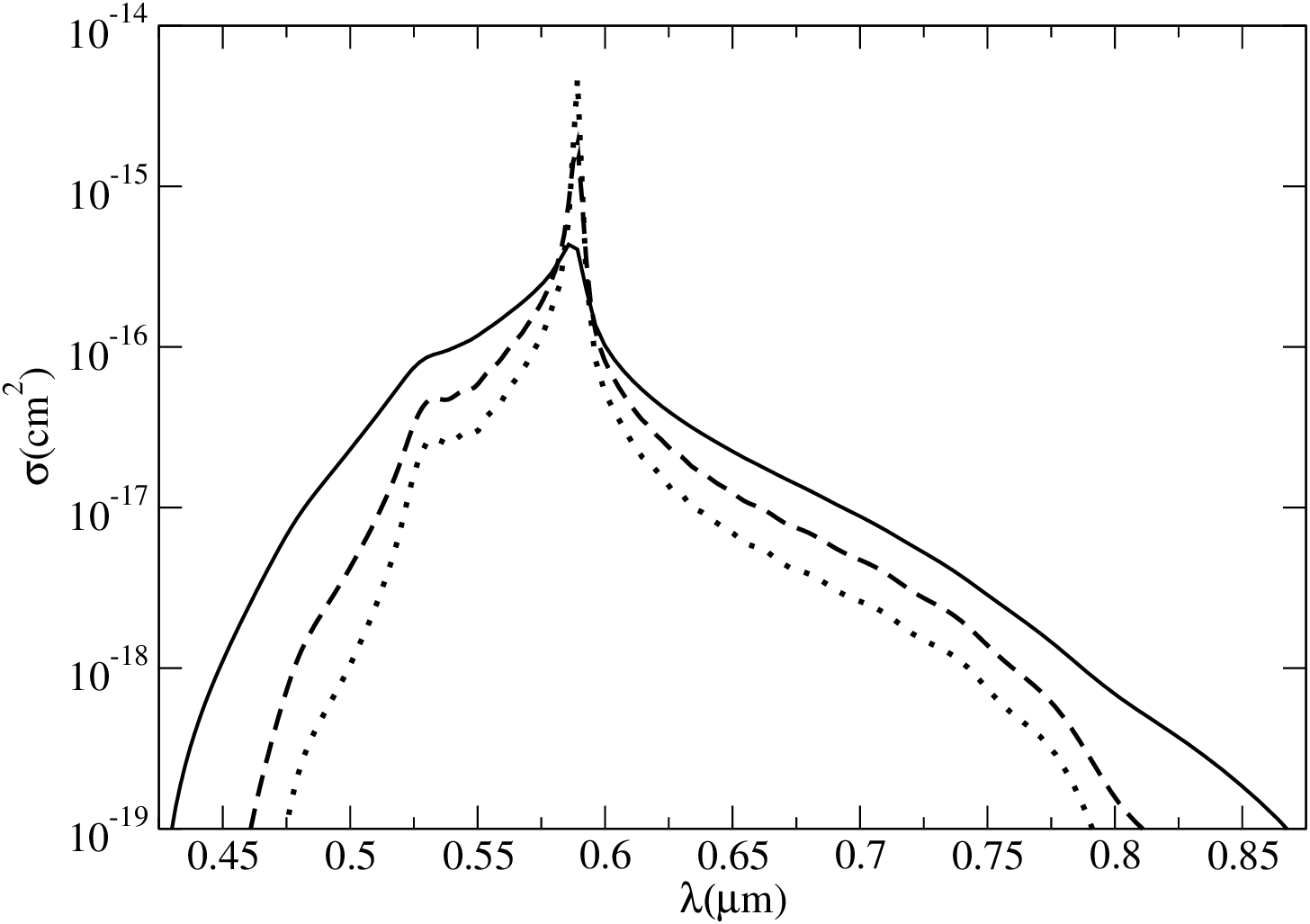}}
\caption  {Variation of the absorption cross section of the $D2$
  component with He density ($T$  = 5000~K,
  $n_{\mathrm{He}}$=5$\times$~$10^{21}$ (solid line),
    2$\times$~$10^{21}$ (dashed line), and $10^{21}$ (dotted line)~cm$^{-3}$).}
\label{vardens}
\end{figure}

At this moderate pressure, the line profile intensities are related to
 the perturbations caused by a single binary collision event. 
However, the density of helium  can reach 
10$^{21}$  cm$^{-3}$ in cool DZ white dwarfs, 
 and for this density multiple perturber effects
 can appear in the blue wing of the $D2$ line (Fig.~\ref{vardens}).
 We can notice the presence of a bump due to  NaHe$_2$ at
about 0.48~$\mu$m,  when the helium density, n$_{He}$ = $10^{21}$ cm$^{-3}$, 
is sufficiently high to observe multiple perturber 
effects \citep{allard2012f}.
 Fig.~\ref{extension} shows the extension of the line wings at
  $T$  = 5000~K and how the unified theoretical approach 
  is a major improvement compared to the unrealistic use of
a  Lorentzian so  far in  the wings.

 The strength  of the line 
satellite at about 0.53~$\mu$m  increases with 
temperature (Fig.~\ref{varT}) and mostly remains at the same position,
 whereas the strength and
 extension of the red wings both increase with  temperature.
There is a total blend of the line satellite in the core of the line
when the density reaches 5$\times$~$10^{21}$ cm$^{-3}$ (Fig.~\ref{vardens}),
leading to the decreasing of the maximum of $\sigma$ of the  $D2$ component.

\begin{figure}[h!]
 \centering
\resizebox{0.46\textwidth}{!}
{\includegraphics*{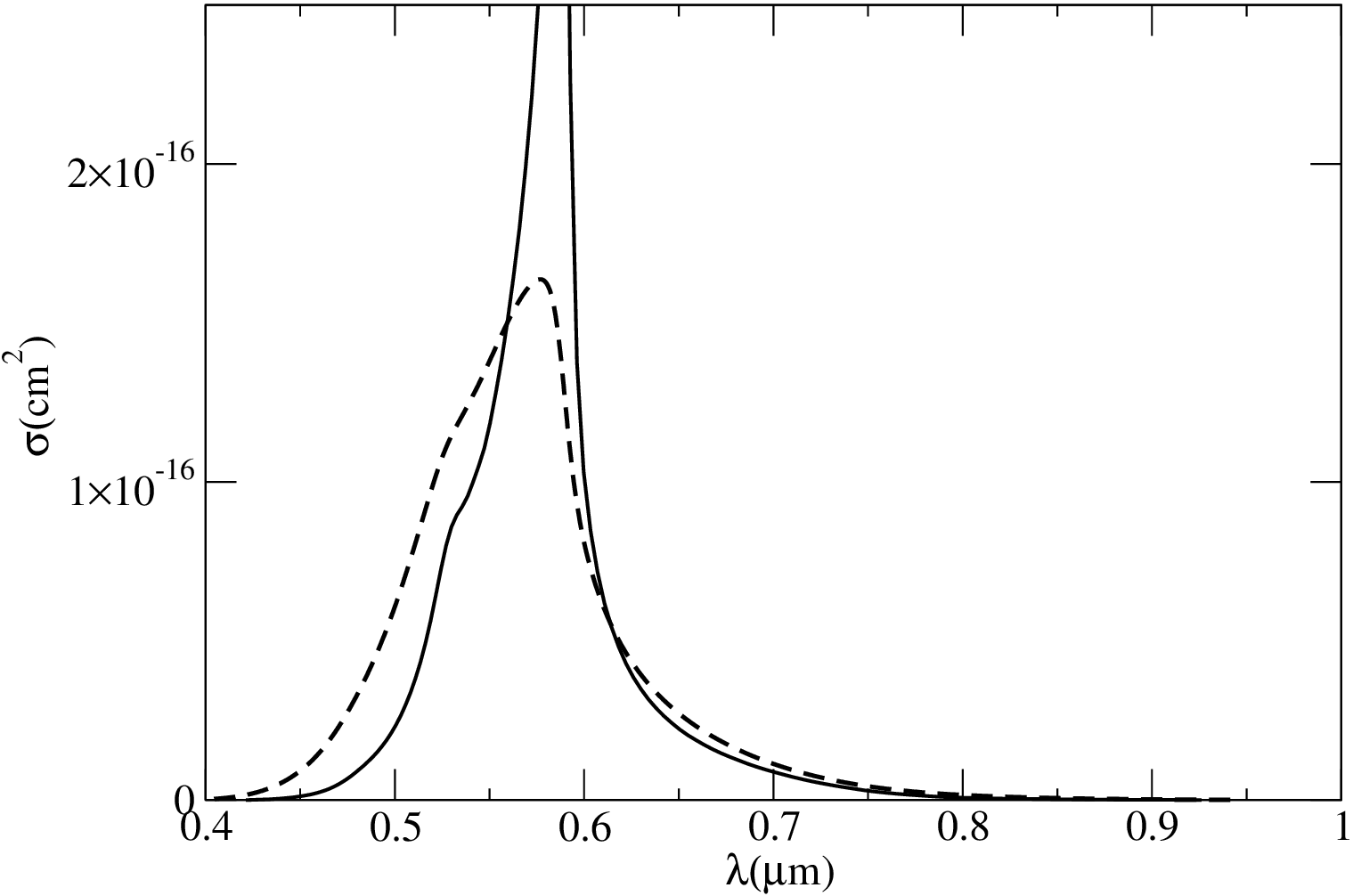}}
\caption{Variation of the absorption cross section of the Na $D2$
 line  with 
helium density ($T$  =5000~K, $n_{\mathrm{He}}$=5$\times$~$10^{21}$  (solid line),
and   $10^{22}$~cm$^{-3}$ (dashed line)).}
\label{nahedens} 
\end{figure}

We note that Na~I line profiles have  been observed in extremely cool,
metal-rich white dwarfs.
The observations of the oldest and colder white dwarfs following the studies
of ~\citet{oppenheimer2001b} and ~\citet{harris2003} have revealed two stars showing
very unusual wide and deep absorption at 5000-6000~\AA\/.
\citet{homeier2007} investigated effects on the Na doublet of high-perturber densities occurring for metal-rich white dwarfs with a
helium-dominated atmosphere. They found that the density of neutral
atomic helium in these two very cool white dwarfs, showing very strong
Na absorption, could reach several 10$^{21}$ to 10$^{22}$ cm$^{-3}$.
Accurate pressure-broadened profiles that are valid at
high densities of He are required to  be incorporated into spectral models.
Previous Na-He opacity tables used by \citet{homeier2007} were constructed
allowing  line profiles up to $n_{\mathrm{He}}$=$10^{19}$ cm$^{-3}$ to be obtained.
Moreover, the line profiles needed to be calculated using  up-to-date
molecular data, which affect the blue
satellite position, the NaHe line satellite is
closer  to the main line than 
 obtained with \citet{pascale1983} (Fig.~\ref{varT}).  
A new  analysis of these stars has  been realized by \citet{blouin2019c} who 
demonstrate that the NaHe satellite  at 0.53~$\mu$m (Fig.~\ref{nahedens}) 
is the large feature observed in the spectrum of WD2356-209.

\begin{figure}
\resizebox{0.46\textwidth}{!}{\includegraphics*{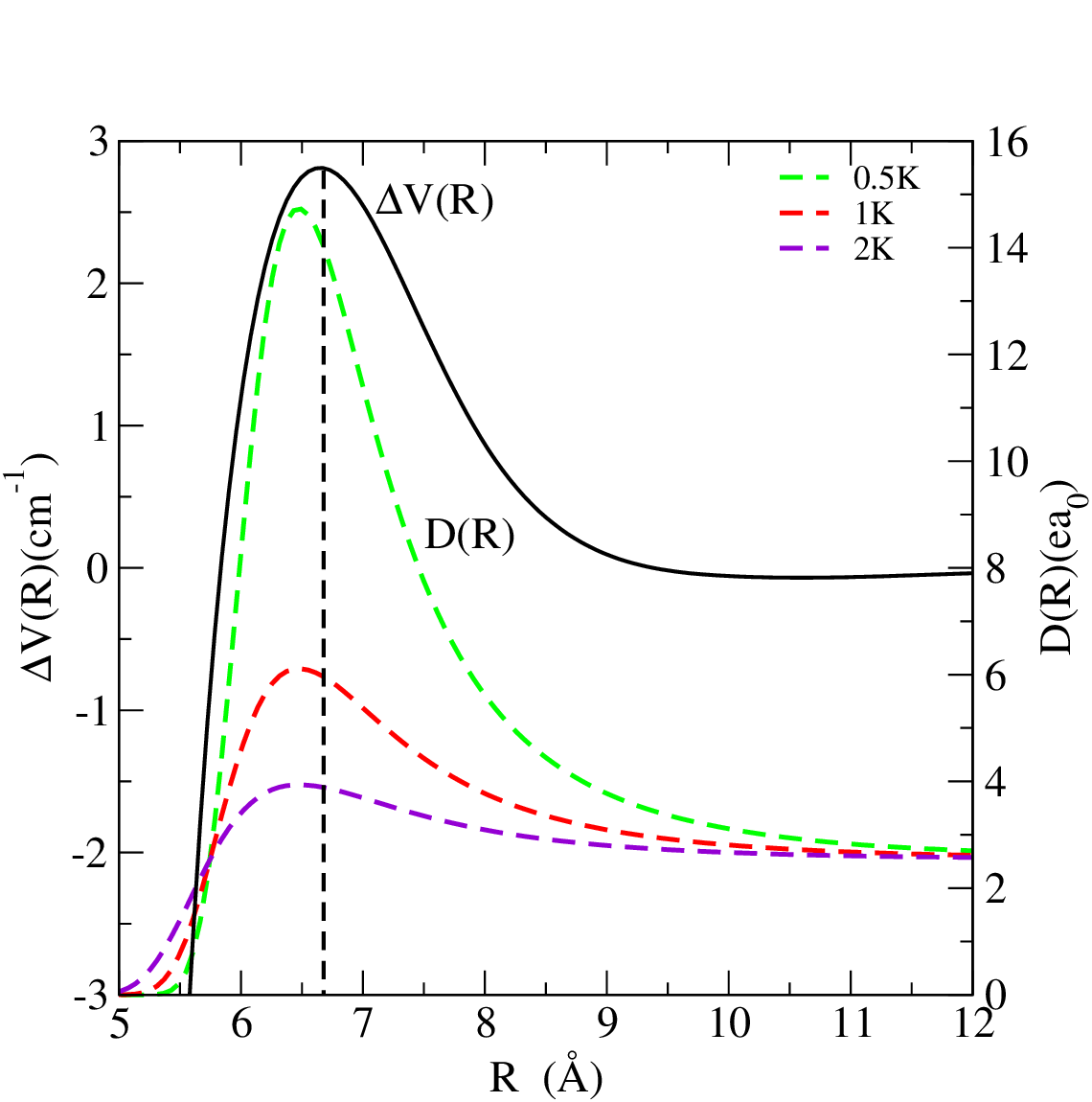}}
\caption  { $\Delta V$  and variation  with  temperatures   of  $D(R)$,
 the Na-He modulated dipole  for the $D$1 component due to the transition 
\mbox {3$s$ $X$  $^2\Sigma^+$   $\rightarrow$ 3$p$ $A$   $^2\Pi_{1/2}$}.
The vertical line is at the maximum of $\Delta V(R)$.
\label{potdipAP12X}}
\end{figure} 

\begin{figure}
\resizebox{0.46\textwidth}{!}
{\includegraphics*{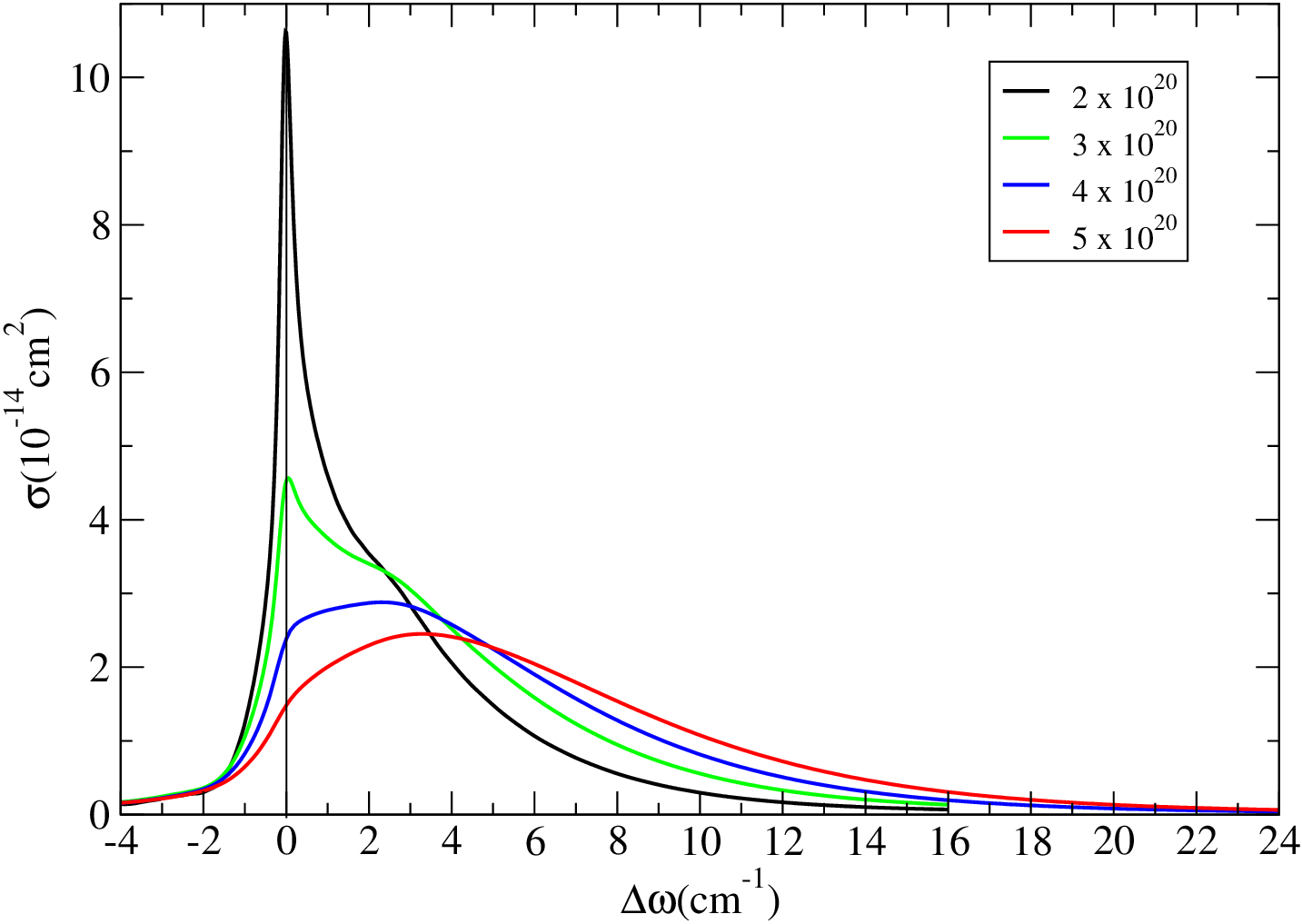}}
\caption {Variation with He atom density of the $D$1 component at T=1~K. We note that
$n_{\mathrm{He}}$ varies from 
 \mbox{2 to 5 $\times $~10$^{20}$cm$^{-3}$} (extracted  from
\citet{allard2013a}).
 \label{nakad1vardens}}
\end{figure}

\subsection{Low temperatures}

At extremely low temperatures, experimental measurements and theoretical
 calculations of helium doped with alkali metals 
have been the subject of active study 
(see~\citep{mateo2011,hernando2010}, and references therein).
Large liquid He clusters produced in a supersonic jet were doped with 
alkali atoms (Li, Na, and K) and characterized by means of laser-induced
fluorescence~\citep{stienkemeier1996}.
The first physical interpretation of the spectra was 
done by~\citet{nakayama2001b}.
Using a diatomics-in-molecules (DIM) type of calculation for 
describing the interaction potential between Na and He atoms,
they performed quantum path integral Monte Carlo  (PIMC) 
simulations to obtain the  He density  profile around 
Na in the ground state and also the absorption spectra of 
Ak-He$_n$ via the Franck-Condon (FC) approximation (Ak=Li, Na, and K).
They employed the semiclassical FC line shape theory for
electronic transitions in a condensed phase system.
In \citet{allard2013a}, we compared our collisional approach 
to PIMC 
calculations of ~\citet{nakayama2001b} using the same  \textit{ab initio} Na-He
molecular potentials. 
Their  Na-He  potentials  and  those of ~\citet{angelo2012}  are superimposed in Figs.~\ref{potx} and \ref{pot3p}, and they are in excellent agreement.
We have shown  in  \citet{allard2013a} that both theoretical approaches 
successfully reproduced the overall shape of the experimental spectra and they
explained distinctive  features through a discussion of the potential difference 
and number of He atoms interacting with the Na atom in the region
of the collision volume.

 At very low temperatures, the multiple perturber effects
 appear in the red wing of the $D1$  component. 
Figure~\ref{potdipAP12X} shows
$D(R)$ together with the corresponding $\Delta V(R)$ 
for  the bound-free transition
 \mbox {3$s$ $X$  $^2\Sigma^+$   $\rightarrow$ 3$p$ $A$ $\Pi_{1/2}$}.
At 1~K, $D(R)$
 is zero for $ R <5 $\AA\/ and is maximum for R=6.4~\AA\/
close to the position of the maximum of $\Delta V$. 
At these low temperatures, $ T <2 $~K, the
long-range potentials appear to be extremely important. The long time
and distance scales 
result in novel features such as quasi-molecular satellites due to the 
 high probability of a bound-free transition.
This increase of $D$(R)  can be explained by the shallow 1.22 cm$^{-1}$ 
well at 6.6~\AA\/ in the ground state potential. 
As shown in Fig.~\ref{potx}, this results from a short range repulsion 
and a weakly attractive interaction at an intermediate range. 

Because $\Delta V$ for  the transition 
 \mbox {3$s$ $X$ $^2\Sigma^+$   $\rightarrow$ 3$p$ $A$ $\Pi_{1/2}$} 
is very small, there is no resolved
peak due to the line satellite; however, the shoulder, at 2.8~cm$^{-1}$ 
in the core of the line, is its signature.
Figure~\ref{nakad1vardens} clearly shows  an asymmetrical
shape of the $D$1 component due to the bound-free transition.
A similar shape was obtained by \citet{allard2016a} for
the triplet $3p$-$4s$ Mg lines perturbed by He,
and the blue asymmetry is a
consequence of low maxima in the corresponding Mg-He potential energy
difference curves at short and intermediate internuclear distances.

Increasing the He density results in a shift of the $D$1 line  toward
the position of the extremum of $\Delta V$ (Fig.~\ref{nakad1vardens}). 
The intensity variation is  evidence of the presence 
of high-order effects, and we observe a blend of these structures.
The complex behavior observed experimentally (Fig.~10 of \citet{allard2013a}) 
reflects the fact that the radiating alkali atom
experiences multiple encounters with He atoms and our collisional
analysis gives a good physical understanding of the phenomena.
In our unified approach, we use averages of independent Na-He pair collision 
processes (Eq.~(\ref{eq:decay})), emphasizing the collisional process
rather than the
collective quantum description ~\citep{nakayama2001b} 
of the interaction of Na with the He
 cluster (which may be important for very small He atom numbers).  
However, our collisional approach goes beyond the FC approximation
and provides a better description close to the line center. 
The main limitation in the simulations obtained 
in the quantum PIMC  theoretical
approach  derives from the fact that the 
FC treatment is valid in the line wing and does not
take into account the line core that is still present when 
red shoulders exist as  in the Na-He absorption spectra 
shown in \citet{stienkemeier1996} and \citet{bunermann2007}.
 Laboratory observations remain a crucial tool for testing the potentials
and assumptions of line shape theory. 
The binding energy of the $X$$^2\Sigma^+$ state and  the interaction energies 
of the excited states at  long range
are close to the limit of the accuracy of  \textit{ab initio} calculations.
In the next section, we examine the dependence of the line parameters on the temperature.

\section{Line core parameters \label{sec:param}}

\subsection{Semiclassical calculations} \label{subsec:sc}

An atomic line broadened by collisions in a low-density gas
has a  Lorentzian profile near the line
center which can be related to the FT
of a radiative wave in which
short duration collisions produce sudden phase changes.
In the theory of an impact-broadened line shape, the phase shifts are given by
Eq.~(\ref{eq:phase})  with the integral  taken between $s=0$ and $\infty$.
The Lorentzian profile can be defined by two line 
parameters, the width and the shift of the main line.
 These quantities can be obtained in the impact limit 
($s$ $\rightarrow$ $\infty$) of the general 
calculation of the autocorrelation function (Eq.~(\ref{eq:gcl})).
In the following discussion, we refer to this line width $w$ as 
measured by half the full width at half the maximum intensity,
what is customarily termed HWHM.

In \citet{allard2007c}, spectral line widths of the light alkalies 
perturbed by He and H$_2$ were 
presented for conditions prevailing in brown dwarf atmospheres. 
We  used pseudo-potentials in a semiclassical
 unified theory of the 
spectral line broadening \citep{allard1999} to compute line core parameters.
For the specific study of the $D1$ (0.5896$\mu\mathrm{m}$) and $D2$ 
(0.5890$\mu\mathrm{m}$) components, we needed to take 
the spin-orbit coupling of the alkali into account.
This was done using an atom-in-molecule intermediate spin-orbit coupling 
scheme, analogous to the one derived by \citet{cohen1974}. 
The asymptotic degeneracy of the $A$$^2\Pi$ and $B$$^2\Sigma$ molecular 
states is partially split by the coupling, resulting in a clear 
energetic distinction between $^2\Pi_{1/2}$ and 
$^2\Pi_{3/2}$ states. We used the molecular-structure calculations
performed by~\citet{pascale1983} for the adiabatic potentials of 
alkali-metal--He systems, and by~\citet{rossi1985} for the molecular 
potentials of alkali--H$_2$ systems.

\begin{figure}
 \centering
\resizebox{0.46\textwidth}{!}
{\includegraphics*{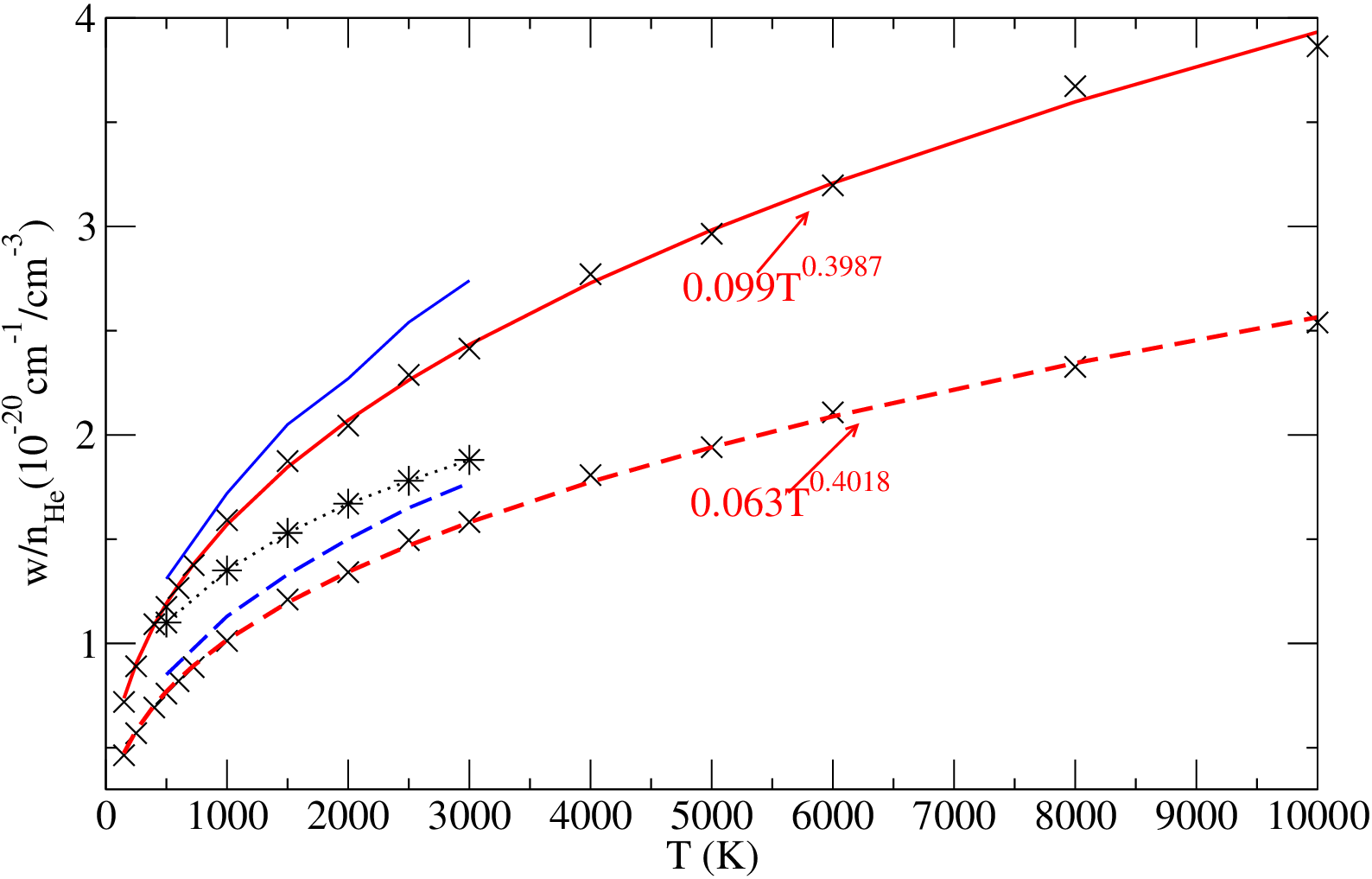}}
\caption  {Variation with temperature of the broadening rate
  ($w$/$n_{\mathrm{He}}$) of the $D2$ (solid line) and
$D1$ (dashed line) resonance lines of Na perturbed by
    He collisions using the potentials of \citet{angelo2012} (red curves),
    \citet{pascale1983} (blue curves), van der Waals potentials (black dotted).
The rates are in units of 10$^{-20}$~cm$^{-1}$/cm$^{-3}$.}
\label{width}
\end{figure}

For He density $n_{\mathrm{He}}$ below  10$^{20}$~cm$^{-3}$,
the core of the line is described adequately by a Lorentzian.
The new determination of line width using \citet{angelo2012} potentials
in a wide range of temperatures  are presented in Figs.~\ref{width} and
Fig.~\ref{param}.
The line widths $w$ (HWHM) are linearly dependent on He density, 
and a power law in temperature is given for the $^2P_{1/2}$ component by
\begin{equation}
w = 0.063 \times 10^{-20} n_{\mathrm{He}} \, T^{0.4018}
\label{eq:wD1vsT}
\end{equation}
and for the $^2P_{3/2}$ component by
\begin{equation}
w = 0.099 \times 10^{-20} n_{\mathrm{He}} \, T^{0.3987}
\label{eq:wD2vsT}
.\end{equation}

These expressions may be used to compute the line widths for temperatures 
of stellar atmospheres from 150 to at least 10000~K.
When it is  assumed that the main interaction between 
two atoms is the long-range van der Waals interaction of 
two dipoles, the Lindholm-Foley theory gives the usual 
formulae for the line width and shift. The van der Waals damping
constant is calculated according to the impact theory of the
collision broadening.

\subsection{Quantum calculations} \label{subsec:qc}

We performed a fully quantum-mechanical calculation of the line core
parameters in order to check their consistency with parameters obtained
from semiclassical theory, presented in Sect.~\ref{subsec:sc}, and to
make a direct comparison with the Na-He quantum calculations of 
\citet{mullamphy2007} and \citet{peach2020}. In the previous section, 
we have shown how the resulting line widths are strongly dependent 
on the Na-He interaction potentials. Our new fully quantum calculations 
of the line core parameters in the  BL 
theory~\citep{baranger1958a} use the \citet{angelo2012} potentials. 
We have noted previously that the potentials used in this work include 
spin-orbit coupling. For direct comparison of the quantum calculations
to the semiclassical calculations of Sect.~\ref{subsec:sc}, which use
the average relative thermal velocity $\bar{v}$, our quantum calculations
use the corresponding relative average wavenumber, $\bar{k}$, to represent
temperature; this quantity is given by
\begin{equation}
\bar{k} = \frac{\mu \bar{v}} {\hbar}
,\end{equation}
where $\mu$ is the reduced mass of the radiating atom and perturber.

Our calculations use the quantum scattering theory of elastic
collisions in a spherically symmetric potential~\citep{merzbacher1998}.
For a given potential energy curve, at each $\bar{k}$ within the 
temperature range $T = 300$--$3000$ K, we solved the radial equations
and computed the scattering phase shifts, $\delta_l(\bar{k})$, for 
partial waves with orbital angular momentum quantum numbers
$l = 0$--$l_{\max}$. The choice of $l_{\max}$ is discussed below.
Phase shifts were computed for potential energy curves of the upper
states, $A\Pi_{1/2}$, $A\Pi_{3/2}$, and $B\Sigma_{1/2}$, and the lower state,
$X\Sigma_{1/2}$.

The BL equations~\citep{baranger1958a} connect the 
scattering phase shifts, $\delta_l(\bar{k})$, to the real and imaginary 
parts of $n_p g(s) \approx (w + i d)s$~\citep{allard1982,peach2020}, which
are the line core width and shift at a given perturber density, $n_p$.
The approximation above is valid when line core asymmetry is negligible.
Convergence criteria for the $w$ and $d$ values make use of the
relation $l_{\max} \approx ka$ for a potential
energy curve of a finite range, $a$~\citep{merzbacher1998}. 
For the Na $D$1 and $D$2 lines, the electronic states perturbed 
by helium all have potential energy curves with nearly the same 
range, $a \approx 30$~a.u. Therefore, for these states we used
\begin{equation}
l_{\max} = 30~\bar{k} + 10
.\end{equation}

In addition to setting $l_{\max}$ for a given state and temperature, 
convergence of the line core parameters also requires the radial
equations to be solved outward to a sufficient internuclear distance,
$R_{\max}$, to obtain accurate partial wave phase shifts in the asymptotic
limit. Examining the convergence of  the broadening and shift rates,
  $w/n_{\mathrm{He}}$ and  $d/n_{\mathrm{He}}$, we found $R_{\max} = 250~a.u.$
  gives  convergence of the rates to the number of significant digits
  shown in Tables B.1 and C.1.

 Quantum calculations of line core parameters are in excellent agreement
with those predicted from semiclassical calculations, over the temperature
range $300$--$3000$~K. Figure~\ref{wd} shows the broadening  and
shift rates versus temperature for the Na-He $D1$ and $D2$ lines,
which were computed using the semiclassical calculation (stars), and using a fully quantum 
calculation (dots). In the absence of performing a Boltzmann average over $k$, 
and instead using $\bar{k}$ at the given temperature, oscillations in
the quantum calculations are evident and expected -- the scattering
of a single coherent plane wave was computed, rather than obtaining a 
weighted sum of coefficients over a distribution of plane waves.  

Broadening and shift rates for Na-He, based on fully quantum 
close-coupling calculations and \textit{ab initio}
potentials, were reported by \citet{mullamphy2007}. 
In Table~\ref{tab:compw} we compare the $D1$ and $D2$ line
broadening rates from our quantum calculations with theirs,
near $T = 500$~K, for which they tabulated the Na-He results.
Table~\ref{tab:compw} also compares the theoretical results with
experimental results. Near $T = 500$~K, we find our theoretical 
$w/n_{\mathrm He}$ for the $D2$ line to be in good agreement with the 
results obtained by \citet{mullamphy2007}, and also in good 
agreement with the experiment. However, our results for the broadening 
rates of the $D1$ line are smaller than those obtained by 
\citet{mullamphy2007} and also with the experiments. We do not 
make a similar comparison of our shift rates, $d/n_{\mathrm He}$, 
for the $D1$ and $D2$ lines with the tabulated values from 
\citet{mullamphy2007} since $d/n_{\mathrm He} \approx 0$ near 
$T = 500$~K. Other previous theoretical and experimental results 
are reported in Table~3 of \citet{allard2007c}.

\citet{mullamphy2007} found their widths to increase as  
$T^{\alpha}$, with $\alpha$ taking the values 0.4389 and 0.3892 
for the $D1$ and $D2$ lines, respectively. The same power-law 
fit to the results of our quantum calculations give $\alpha$ to be 
$0.4116\pm 0.0111$ and $0.3976\pm 0.0041$. Fitted values of 
$\alpha$ agree, to within the fitting uncertainties, with 
those obtained from our semiclassical calculations and presented 
in sec.~\ref{subsec:sc}.

\begin{table}
\caption{Comparison of experimental and theoretical broadening
rates $w/n_{\mathrm He}$ ($10^{-20}$~cm$^{-1}$/cm$^{-3}$) of 
the Na resonance lines.}
\label{tab:compw}
\begin{tabular}{lccc}
  \hline
\noalign{\medskip}  
\multicolumn{1}{c}{Reference} &
\multicolumn{1}{c}{$D1$} &
\multicolumn{1}{c}{$D2$} &
\multicolumn{1}{l}{T(K)} \\
\hline
\noalign{\medskip}
        {This work (BL)} & 0.7127 & 1.123 & 450 \\
\noalign{\medskip}        
\citet{mullamphy2007} & 0.9507 & 1.109 & 450 \\
\noalign{\medskip}
\citet{kielkopf1980} & {$1.02\pm 0.01$} & {$1.06\pm 0.02$} & {$450\pm 50$} \\
\hline
\noalign{\medskip}
        {This work (BL)} & 0.7314 & 1.1548 & 475 \\
\noalign{\medskip}        
\citet{mccartan1976} & {$1.01\pm 0.05$} & {$1.16\pm 0.10$} & {$475\pm 15$} \\
\hline
\noalign{\medskip}
        {This work (BL)} & 0.7350 & 1.1612 & 480 \\
\noalign{\medskip}        
\citet{mullamphy2007} & 0.9797 & 1.138 & 480 \\
\hline
\end{tabular}
\end{table}

\subsection{Contribution of the different components}

In Tables~B1 and C1, we report our computed  broadening rates
and shift rates obtained in the semiclassical theory compared to the quantum calculation for the different transitions. Figure~\ref{param} shows their dependence on temperature.
The $D1$ line is due to a simple isolated $A$ $\Pi_{1/2}$ state,
whereas the $D2$ line comes from the $A$ $\Pi_{3/2}$ and 
$B$ $\Sigma_{1/2}$ adiabatic states arising from the $3p$ $P_{3/2}$ atomic 
state.  The broadening of the $B$ $\Sigma_{1/2}$--$X$ $\Sigma_{1/2}$
transition is most sensitive to potential at intermediate to short-range 
separations ($R < 5$~\AA). This result confirms the study by 
\citet{roueff1969} and \citet{lortet1969} of the collisions with 
light atoms whose polarizability is small. It was shown by them that 
the width of spectral lines due to collisions with hydrogen atoms 
does not arise from the van der Waals dispersion forces,
but from a shorter range interaction.

\begin{figure}
 \centering
\resizebox{0.46\textwidth}{!}
{\includegraphics*{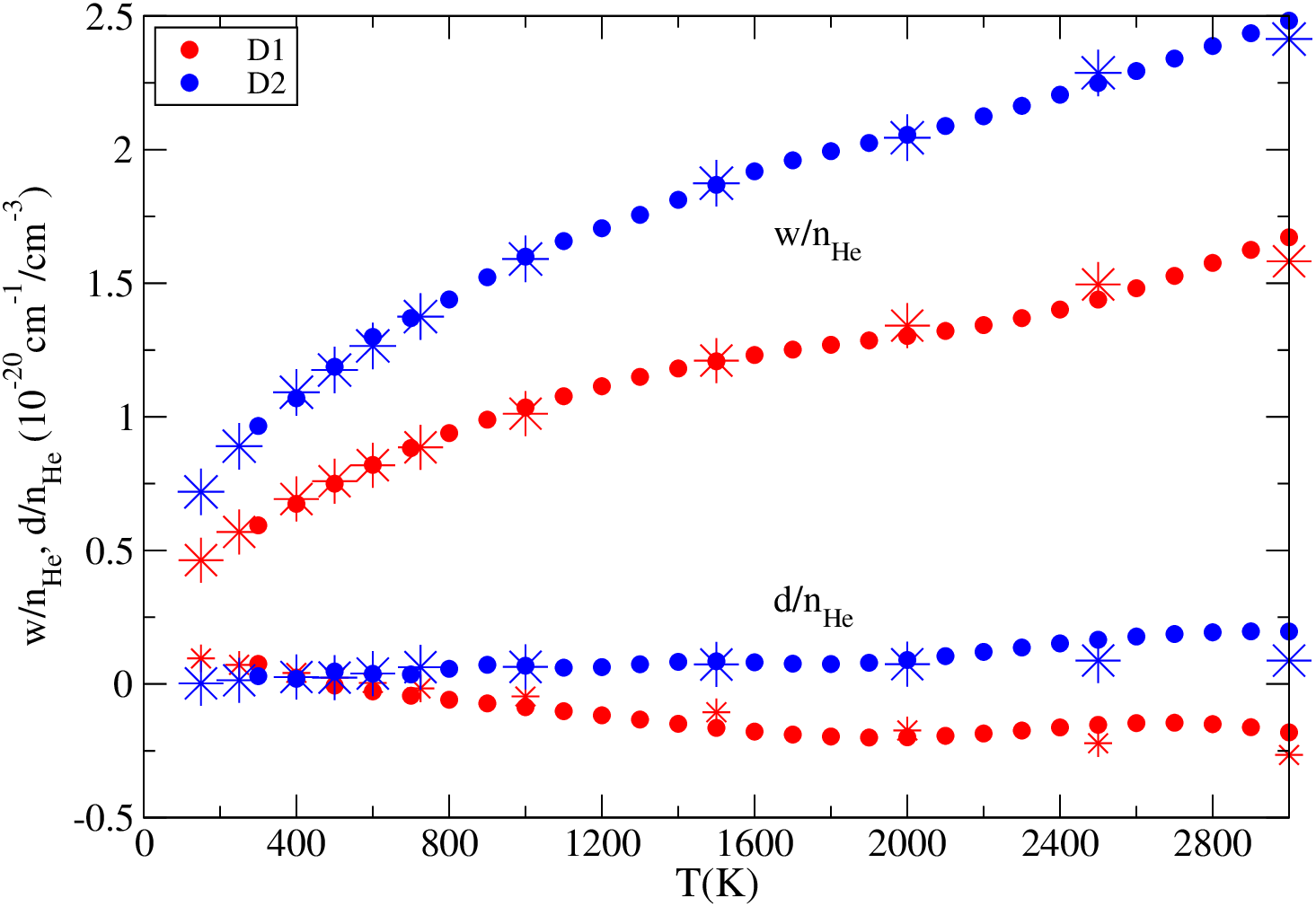}}
\caption  {Variation with temperature of the
  broadening rate  ($w$/$n_{\mathrm{He}}$) and of the
  shift rate ($d$/$n_{\mathrm{He}}$) of the $D1$ (red) 
  and $D2$ (blue) resonance lines of Na perturbed by He 
  collisions: BL(circles), semiclassical (stars).  
The rates are in units of 10$^{-20}$~cm$^{-1}$/cm$^{-3}$.}
\label{wd}
\end{figure}

\begin{figure}
 \centering
\resizebox{0.46\textwidth}{!}
{\includegraphics*{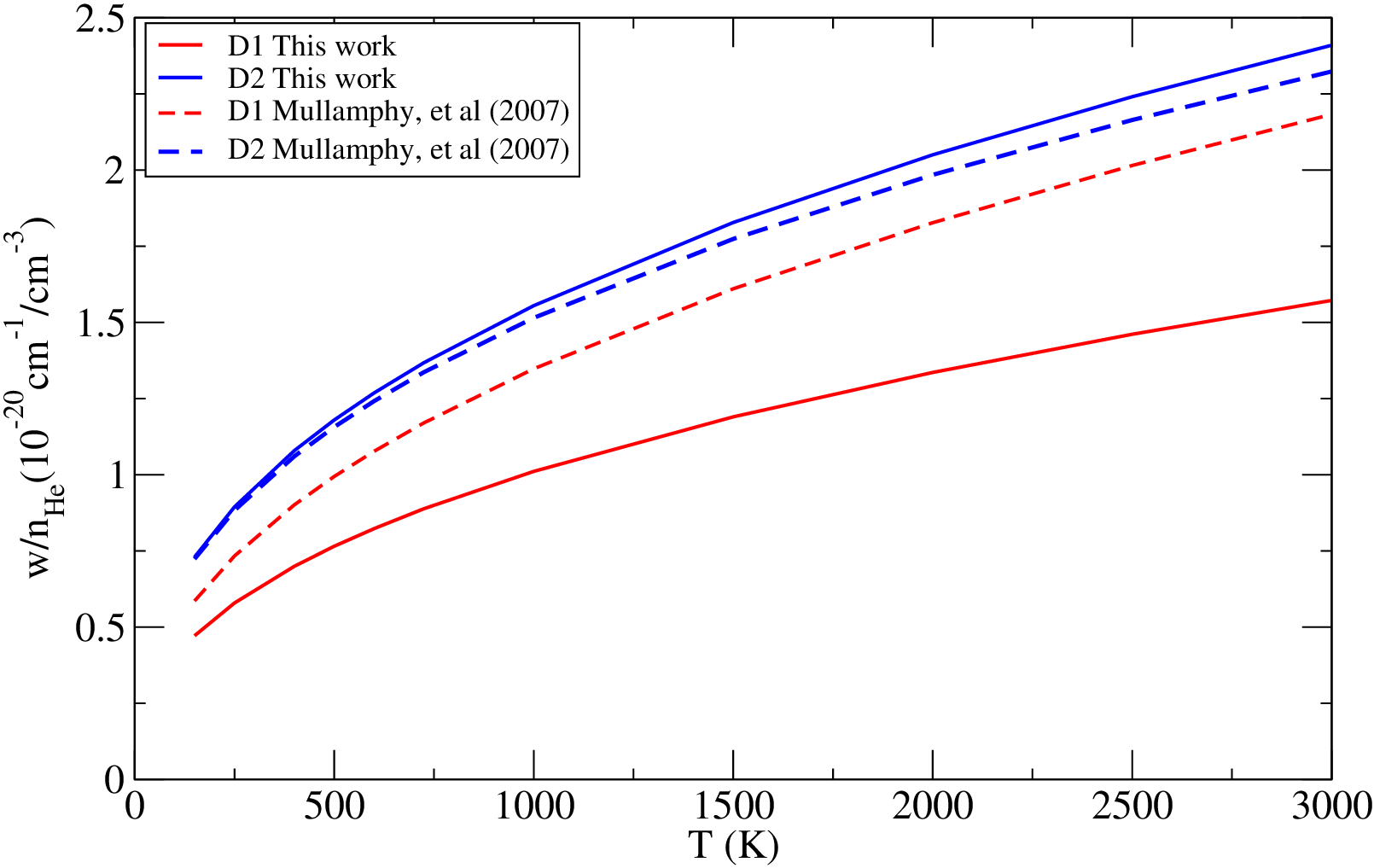}}
\caption{Comparison of theoretical broadening rates ($w$/$n_{\mathrm{He}}$) 
for the $D1$ (red) and $D2$ (blue) resonance lines, using the power-law fitted 
temperature dependence from the present work (solid), 
Eqs.\ref{eq:wD1vsT}--\ref{eq:wD2vsT},
with the power laws obtained by \citet{mullamphy2007} (dashed).
The rates are in units of 10$^{-20}$ cm $^{-1}$/cm$^{-3}$.}
\label{wcomp}
\end{figure}

The combined contribution of the $A$ $\Pi_{3/2}$--$X$ $\Sigma_{1/2}$
and $B$ $\Sigma_{1/2}$--$X$ $\Sigma_{1/2}$ transitions leads to a larger 
width for the $D2$ line, as seen in Fig.~\ref{wd}.  We may compare 
the results of our \textit{ab initio} calculations for the broadening rate 
versus temperature of the $D1$ and $D2$ lines with those obtained by 
\citet{mullamphy2007}. The power-law fits for our semiclassical and 
quantum calculations were seen to agree to within the fitting 
uncertainties. Therefore, we compared the broadening rates from 
Eqns.~\ref{eq:wD1vsT} and \ref{eq:wD2vsT} with the corresponding 
power-law fits obtained by \citet{mullamphy2007} and given, in their 
Table~4, for the $D1$ and $D2$ resonance lines of Na perturbed by 
He. The comparison of the theoretical broadening rates, shown in 
Fig~\ref{wcomp}, reveals that, at higher temperature, our $D2$ 
line-broadening rate is slightly higher than the result obtained 
by \citet{mullamphy2007}, while our $D1$ line-broadening 
rate is significantly lower than the \citet{mullamphy2007}
result.

\begin{figure}
 \centering
\resizebox{0.46\textwidth}{!}
{\includegraphics*{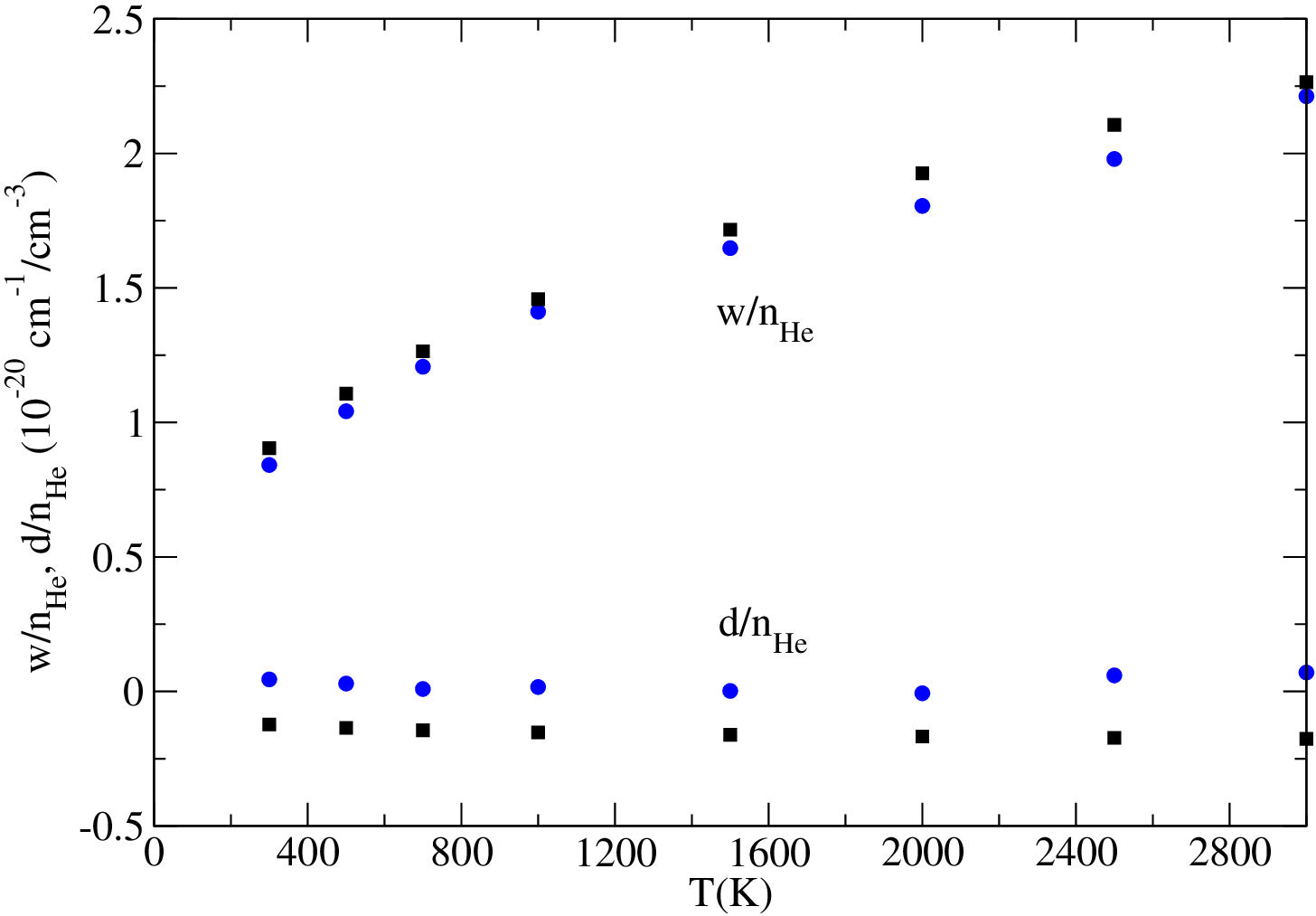}}
\caption  {Comparison of the
  broadening rate  ($w$/$n_{\mathrm{He}}$) and of the
  shift rate ($d$/$n_{\mathrm{He}}$) for the unresolved 
  $3P\mathrm{-}3S$ doublet according to quantum calculations of the current work (blue circles) and as reported by \citet{peach2020} (dark squares).  
The rates are in units of 10$^{-20}$ cm $^{-1}$/cm$^{-3}$.}
\label{wdcomp}
\end{figure}

We may also compare our quantum calculations of broadening and shift
rates, shown in Fig~\ref{wd}, with those of \citet{peach2020} by combining
the $D1$ and $D2$ rates at a given temperature into a single rate for
$3P$-$3S$, since \citet{peach2020} give the rates versus temperature only 
for the unresolved doublet. The appropriate combination of $D1$ and $D2$
rates, based on the relative line strengths for the two transitions, is
\begin{equation}
r(3P\mathrm{-}3S) = 2r(D2) + r(D1)
,\end{equation}
where $r$ is the rate $w/n_p$ or $d/n_p$. Figure~\ref{wdcomp} compares 
our quantum calculated rates with those of \citet{peach2020} at selected
temperatures. The differences between our broadening rates for the 
$D1$ and $D2$ lines and those of \citet{mullamphy2007}, 
shown in Table~\ref{tab:compw} and Fig~\ref{wcomp}, and between our 
broadening and shift rates for the unresolved doublet with those of 
\citet{peach2020}, shown in Fig.~\ref{wdcomp}, are attributed to differences 
in the potential energy curves used in the respective calculations.

\section{Conclusions}

Ultracool stellar atmospheres show absorption by alkali resonance
lines severely broadened by collisions with neutral perturbers.
 In the coolest and densest
  atmospheres, such as those of T dwarfs, Na I and K I broadened by
  molecular hydrogen and helium can come to dominate the entire
  optical spectrum.  The effects of collision broadening by He are also
  central to understanding the opacity of cool DZ white dwarf stars
 which have high-pressure atmospheres. We have carried out a detailed analysis of the
Na-He line profiles  used for  astrophysical applications for  cool DZ 
white dwarf spectra which show  extreme absorption by Na.
 In order to be able to construct synthetic spectra of brown dwarfs and cool DZ
  white dwarfs, when helium density can become as high as 10$^{21}$
  cm$^{-3}$, line profiles were computed in a wide range of
  densities and temperatures.
All line shape characteristics depend on
 the accuracy of the   \textit{ab initio} potentials and of the
 suitable line shape theory.
  As part of a more
general or unified concept of spectral line formation, quasi-molecular 
line satellites, which appear in the line wing of alkali resonance lines 
perturbed by helium or molecular hydrogen, are the
binary-collision manifestation of a ubiquitous phenomenon, the many-body nature
of spectral lines from dense gases.
Collision-broadened lines show features that arise from the changed structure of
the radiating atom in the presence of another atom.  As such, their presence is
sensitive to the conditions of the radiating system and reveal the composition,
temperature, and pressure of the source.

New Na--He opacity tables were constructed. The complete tables are available on request. They will be archived  at the CDS, 
 with an explanation of their use and a program to produce
 line profiles for the $D1$ and $D2$ components
 to $n_{\rm He}$=$10^{21}$ cm$^{-3}$ 
 from $T_\mathrm{eff}=150\ \mathrm{K}$ to 2500~$\ \mathrm{K}$ and to
 $n_{\rm He}$=$10^{22}$ cm$^{-3}$ 
 from $T_\mathrm{eff}=3000\ \mathrm{K}$ to 10 000~$\ \mathrm{K}$.

\begin{acknowledgements}
   We thank the referee for insightful comments.
\end{acknowledgements}


\begin{appendix}

  \onecolumn

\section{Multiple perturber effects: Formation of
    NaHe$_n$ quasi-molecular satellites}
  
\begin{figure}[h!]
 \centering
\resizebox{0.46\textwidth}{!}
{\includegraphics*{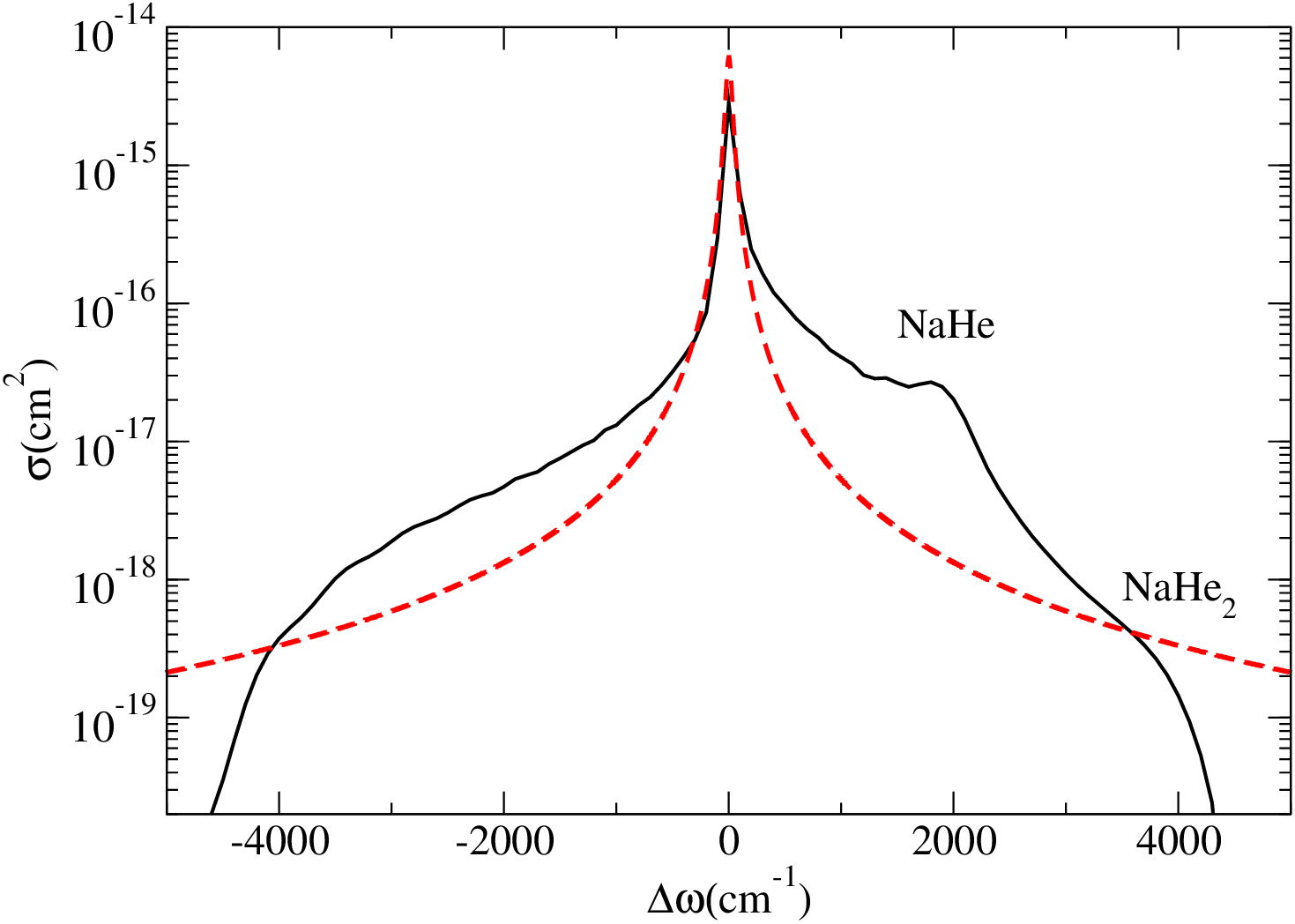}}
\caption{Absorption cross section of the Na $D2$ component at 
 $T$  =5000~K, $n_{\mathrm{He}}$=1$\times$~$10^{21}$  (solid line) 
compared to the corresponding Lorentzian profile (dashed line)).}
\label{extension} 
\end{figure}

\begin{center}

\section{Line broadening rate}

{Table B.1. Computed broadening rates $w/n_{\mathrm{He}}$ 
 (10$^{-20}$ cm$^{-1}$/cm$^{-3}$)
  of Na resonance lines perturbed by He collisions. Values are given from
  both semiclassical (sc) and quantum (BL) calculations.}
\label{tab:wNaHe}
\begin{flushleft}
\begin{tabular}{llllllllll}
\noalign{\bigskip}
\hline
\noalign{\bigskip}
\multicolumn{1}{l}{Component} &
\multicolumn{1}{l}{Transition} &
\multicolumn{1}{l}{Calc.} &
\multicolumn{1}{l}{500~K} &  
\multicolumn{1}{l}{1000~K} &
\multicolumn{1}{l}{1500~K} & 
\multicolumn{1}{l}{2000~K} & 
\multicolumn{1}{l}{2500~K} & 
\multicolumn{1}{l}{3000~K} &
\multicolumn{1}{l}{weight}\\
\noalign{\bigskip}
\hline
\noalign{\bigskip}
 $3s$ $^2S_{1/2}$-$3p$ $^2P_{1/2}$ & $A$ $\Pi_{1/2}$-$X$ & sc
& 0.759 & 1.012  & 1.21 & 1.34 & 1.50 & 1.58 & 1\\
\noalign{\medskip}
\noalign{\medskip}
  && BL
 & 0.750   &   1.035 & 1.208 & 1.303 & 1.439  & 1.672 & 1 \\
\noalign{\medskip}
\noalign{\medskip}
  $3s$ $^2S_{1/2}$-$3p$ $^2P_{3/2}$  &  $A$ $\Pi_{3/2}$-$X$ & sc
& 0.426 & 0.557 & 0.659 & 0.731 & 0.809 & 0.854  & 0.5\\
\noalign{\medskip}
\noalign{\medskip}
  && BL
 & 0.432  & 0.578 & 0.670 & 0.717 & 0.772 & 0.886 & 0.5\\
\noalign{\medskip}
\noalign{\medskip}
   & $B$ $\Sigma_{1/2}$-$X$ & sc
 & 0.75 & 1.034 & 1.216   & 1.314 & 1.479 & 1.56 & 0.5  \\
\noalign{\medskip}
\noalign{\medskip}
&  & BL
& 0.755 & 1.022 & 1.198  & 1.339 & 1.478  & 1.597   & 0.5 \\
\noalign{\medskip}
\noalign{\medskip}
&  $^2P_{3/2}$ & sc
& 1.175 & 1.591 & 1.874 & 2.045 &  2.288 & 2.414 & \\
\noalign{\medskip}
\noalign{\medskip}
&& BL
 & 1.187 & 1.600 & 1.868 & 2.057 & 2.249 & 2.483 & \\
\noalign{\medskip}
\noalign{\medskip}
\hline
\end{tabular}
\\
\end{flushleft}

\newpage

\section{Line shift rate}

{Table C.1. Computed shift rates $d/n_{\mathrm{He}}$
 (10$^{-20}$ cm$^{-1}$/cm$^{-3}$)
  of Na resonance lines perturbed by He collisions. Values are given from
  both semiclassical (sc) and quantum (BL) calculations.}
\label{tab:dNaHe}
\begin{flushleft}
\begin{tabular}{llllllllll}
\noalign{\bigskip}
\hline
\noalign{\bigskip}
\multicolumn{1}{l}{Component} &
\multicolumn{1}{l}{Transition} &
\multicolumn{1}{l}{Calc.} &
\multicolumn{1}{l}{500~K} &  
\multicolumn{1}{l}{1000~K} &
\multicolumn{1}{l}{1500~K} & 
\multicolumn{1}{l}{2000~K} & 
\multicolumn{1}{l}{2500~K} & 
\multicolumn{1}{l}{3000~K} &
\multicolumn{1}{l}{weight}\\
\noalign{\bigskip}
\hline
\noalign{\bigskip}
 $3s$ $^2S_{1/2}$-$3p$ $^2P_{1/2}$ &  $A$ $\Pi_{1/2}$-$X$ & sc
& 0.023  & -0.047  & -0.106  & -0.174  & -0.222 & -0.266  & 1\\
\noalign{\medskip}
\noalign{\medskip}
  && BL
& -0.006  &   -0.087 & -0.164 & -0.199 & -0.153  & -0.181 & 1 \\
\noalign{\medskip}
\noalign{\medskip}
  $3s$ $^2S_{1/2}$-$3p$ $^2P_{3/2}$  & $A$ $\Pi_{3/2}$-$X$ & sc
& -0.147 & -0.196 & -0.231 & -0.274  & -0.298  & -0.329  & 0.5\\
\noalign{\medskip}
\noalign{\medskip}
& & BL
& -0.162  & -0.210 & -0.260 & -0.293 & -0.275 & -0.276 & 0.5\\
\noalign{\medskip}
\noalign{\medskip}
   & $B$ $\Sigma_{1/2}$-$X$ & sc
 & 0.171 & 0.260 & 0.303 & 0.348  & 0.386 & 0.416 & 0.5  \\
\noalign{\medskip}
\noalign{\medskip}
&  & BL
& 0.209 & 0.278 & 0.345  & 0.383 & 0.441  & 0.472  & 0.5 \\
\noalign{\medskip}
\noalign{\medskip}
&  $^2P_{3/2}$ & sc
& 0.024 & 0.064 & 0.073 & 0.074 & 0.088 & 0.088 & \\
\noalign{\medskip}
\noalign{\medskip}
&& BL
& 0.047 & 0.068 & 0.085 & 0.090 & 0.166 & 0.196 & \\
\noalign{\medskip}
\noalign{\medskip}
\hline
\end{tabular}
\\
\end{flushleft}

\end{center}

\section{Line parameters}

\begin{figure}[h!]
 \centering
\resizebox{0.46\textwidth}{!}
{\includegraphics*{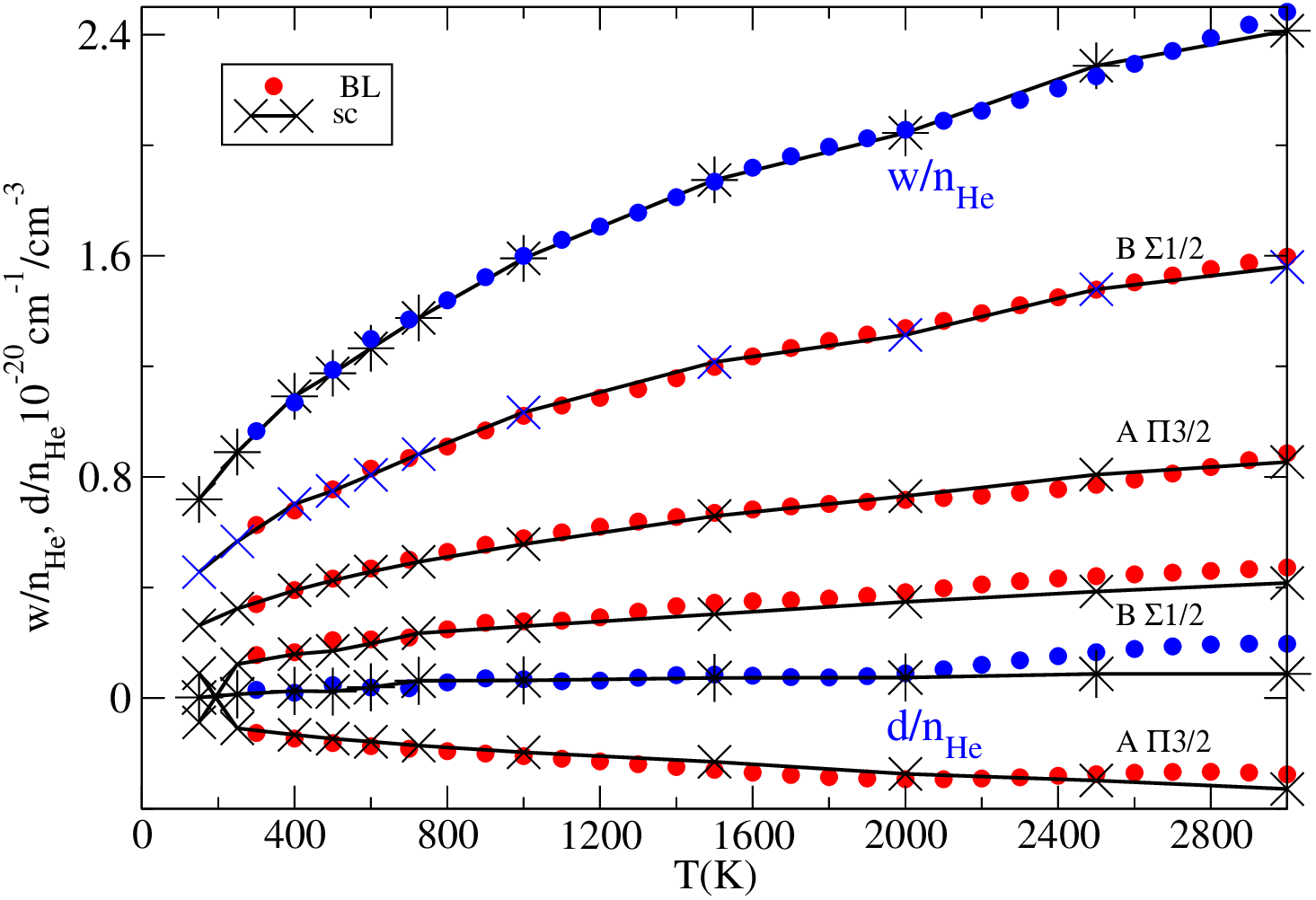}}
\caption  {Comparison  of the broadening rate
  ($w$/$n_{\mathrm{He}}$) and of the shift rate ($d$/$n_{\mathrm{He}}$)
  of the $D2$ line compared to the contribution of the
  different transitions  $B$ $\Sigma_{1/2}$-$X$ and $A$ $\Pi_{3/2}$-$X$:
  BL (circles) and semiclassical (stars).  
The rates are in units of 10$^{-20}$ cm $^{-1}$/cm$^{-3}$.}
\label{param}
\end{figure} 

\end{appendix}

\end{document}